\documentstyle[12pt]{article}
\def\lromn#1{\uppercase\expandafter{\romannumeral#1}}

\def\blist{\begin{list}{\setlength{\rightmargin}{\leftmargin}}}
\def\elist{\end{list}}

\begin{document}

\begin{titlepage}

\begin{flushright}
June 1995

TU/95/484\\
hep-th/9506176
\end{flushright}

\vspace{12pt}

\begin{center}
\begin{Large}

\renewcommand{\thefootnote}{\fnsymbol{footnote}}
\bf{
Catastrophic Particle Production \\
under Periodic Perturbation
}
$^{*}$
\end{Large}

\vspace{36pt}

\begin{large}
\renewcommand{\thefootnote}{\fnsymbol{footnote}}
M.Yoshimura
$^{**}$

Department of Physics, Tohoku University\\
Sendai 980 Japan\\
\end{large}

\vspace{54pt}

{\bf ABSTRACT}
\end{center}

\vspace{0.5cm}
We develop a formalism to investigate the behavior of quantum field and
quantum ground state when the field is coupled
to perturbation that periodically oscillates.
Working in the Schr$\stackrel{..}{{\rm o}}$dinger picture of quantum field
theory, we confirm
that the phenomenon of parametric resonance in the classical theory
implies an instability of quantum vacuum,
and correspondingly it gives rise to catastrophic particle
production if the oscillation lasts indefinitely;
the produced number of particles exponentially
increases without bound as time proceeds.
The density matrix describing the limiting stage of the quantum state
is determined by a small set of parameters.
Moreover, the energy spectrum and the intensity of produced particles
are worked out in greatest detail in the limit of weak coupling or
small amplitude perturbation.
In the case of strong coupling or large amplitude perturbation
the leading adiabatic formula is derived.
Application to cosmological fate of weakly interacting spinless
fields (WISF) such as the invisible axion, the Polonyi, and the modular fields
is discussed.
Although very little effect is expected on the invisible axion, the Polonyi
type field has a chance that it catastrophically decays at an early
epoch without much production of entropy, provided that an intrinsic
coupling is large enough.

\end{titlepage}

{\bf
\lromn 1 \hspace{0.2cm}
Introduction
}

\vspace{0.5cm}
What would be the fate of the quantum ground state if a periodic oscillation of
some perturbation lasts for an indefinitely long time?
This seemed a purely accademic problem, until modern unified theories of
the micro-world predicted many weakly interacting spinless fields (WISF),
and until it was clarified
that condensate oscillation of these fields may become the major
massive component of our universe,
at least during some epoch of the early universe.
Indeed, in some case (the invisible axion case \cite{axioncosmo})
this type of WISF becomes the cold dark matter candidate,
even of the present universe, while in another case
(the Polonyi field case \cite{polonyi})
their decay creates too huge an entropy,
destroying either the standard scenario of nucleo-synthesis or
baryo-genesis,
which then eliminates an interesting class of supergravity models.

In this paper we shall formulate the fundamental problem of attacking
the behavior of quantum field and quantum state under periodic
perturbation.
Our formalism is based on the Schr$\stackrel{..}{{\rm o}}$dinger picture of
quantum field theory. In this formulation relationship between the
quantum wave function and the classical oscillator motion
becomes evidently clear. Thus we are able to
confirm without any ambiguity that the exponential decay law
of the initial ground state is related to the classical instability
(parametric resonance) under periodic perturbation.
Moreover, we are able to work out the density matrix of
the quantum  state initially started as the ground state,
and able to compute the energy spectrum and
the intensity of produced particles at any asymptotic future time.

The phenomenon of catastrophic particle production that occurs at
the limiting late stage of periodic perturbation has many features so
distinct as to be verified unmistakably if it ever can be checked by
laboratory experiment.
It may also have important implication on cosmological WISF.
We briefly examine the two cases of immediate interest,
the invisible axion and the Polonyi field.
It is shown that although very little effect is expected for the axion,
the Polonyi type field has a chance of being dissipated away
without much entropy creation.
This would be a blessing to some phenomenology of supergravity models, but
we leave to future the final word on this issue until more details are
worked out.

Some aspect of particle production under periodic perturbation
has been investigated in the past, especially
in regard to the reheating problem after inflation
\cite{reheating parametric}, \cite{reheating recent}.
Aside from our different motivation on WISF, we point out that
our main achievement in the basic formalism
is to clarify the nature of quantum state at the level of the density matrix,
which has not been available so far even in the limited situation.
Only with this level of understanding one can completely
determine the limiting
behavior of the quantum ground state, and can compute many quantities
from the first principle.
In clarifying the precise nature of quantum state, it is crucial to
implement the unitarity, namely conservation of the probability.
Our formulation manifestly satisfies the unitarity at every stage of
approximation, whereas in some calculation in the literature the unitarity is
obscured.
We hope that we have contributed much to this regard.
In technical aspects, we mention two major improvements here over previous
works:
explicit formulas from all higher resonance bands in the small amplitude
or the weak coupling case, and derivation of a major adiabatic contribution
in the large amplitude or the strong coupling case.

The rest of this paper is organized as follows.
In explaining the content of each section, we shall also state our main
conclusion so that the reader can regard this as a summary of our work.
In Section \lromn 2 we formulate the Schr$\stackrel{..}{{\rm o}}$dinger
picture of the quantum ground state subjected to continuous
periodic perturbation after an initial time.
We only consider the simplest type of model of this sort, the Gaussian
model with a periodic mass function.
The wave functional of the entire system can be decomposed into a product of
wave functions of  Fourier modes since spatial translational invariance
is kept intact. The Gaussian wave function of each mode is then
written in terms of a single function of time $u(t)$ that obeys the classical
motion of field oscillator under the periodic potential. The width of the
Gaussian peak is given by the modulus $|u(t)|$, and the phase in
the Gaussian exponent by the logarithmic derivative of the modulus.
Thus analysis of the quantum wave function is reduced to that of
the classical equation under some specific initial condition, determined
from the initial condition of the quantum system.
The spread of the Gaussian peak
corresponds to excitation of higher levels of harmonic oscillator,
thus production of particles. Hence growth of the field oscillator
amplitude in the classical equation
implies an increase of particle production in the quantum state.

In Section \lromn 3 we explain how general Gaussian model is described
by the density matrix in such a way that the unitarity is manifest.
The density matrix in the Fock base is much easier, in the Gaussian model,
to handle than the wave function itself, and is
much more convenient to compute various observable quantities.
The diagonal element of the density matrix, namely the probability
distribution of $n$ particle production,
is written in terms of a quantity that may be interpreted as
the average number of produced particles.

Section \lromn 4 is a mathematical addendum in which we first explain
how the instability region of the parametric resonance may
be identified and derive the width of the resonance band in the case of small
amplitude perturbation. Our method also
makes it possible to compute the growing mode function in a systematic
series expansion of the coupling factor.
We then sketch how to obtain
solution to the same classical equation under the large
amplitude perturbation.
The adiabatic approximation is excellent in this regime.
The classical amplitude of field oscillator is given in this case by
an elliptic integral of the second kind, from which we determine
the limiting behavior.
This section contains some useful formulas and new derivation of old results
that are not written in
the standard textbook \cite{mathieu eq} on the Mathieu equation.
Detailed derivation of weak coupling formulas is relegated to Appendix.

In Section \lromn 5 we combine results of the previous sections
in order to determine
the limiting behavior of the evolved ground state.
Under the small amplitude perturbation
there exist two vastly different time scales in the problem,
the short time scale of oscillation, and the long time scale of decay
or the inverse of the production rate of particles.
We first time-average over the short time scale, ignoring transient phenomena.
We then mode-sum over frequencies within the resonance band.
In this way it becomes possible to obtain a useful formula of the
spectrum intensity at any asymptotic time.
Adding contribution from resonance frequencies in
infinitely many bands gives rise to
the exponential decay law of the ground state and its associated particle
production. A precise relation between the two is established.
If the periodic perturbation lasts indefinitely, the particle production
becomes catastrophic.
In the limit of small amplitude perturbation
the spectrum of produced particles is equally spaced at
$\omega = \frac{E}{2}\,n \; (n = 1 \,, 2\,, 3 \cdots )$
of increasingly narrower band width, and
the intensity of each band is determined as a function of time.
In the large amplitude regime the particle production rate has a dependence
on the oscillation amplitude quite different from the small amplitude
case. This is shown by using the leading adiabatic formula in the previous
section.

Final Section \lromn 6 describes cosmological application on WISF.
We first explain briefly the cosmological damped oscillation of WISF.
After the onset of the oscillation the Hubble time is  time
scale of the amplitude damping, hence during one Hubble time one may ignore
effect of damping and assume undamped periodic oscillation.
After introducing a general parametrization on WISF properties,
we discuss effect of the cosmological redshift that may push away
relevant frequencies outside the resonance band.
Rate of particle production while staying
within the resonance band is a real measure of the decay rate of
cosmologial WISF.
It turns out
that frequencies in general redshift away in time much shorter than
one Hubble time under the small amplitude perturbation of WISF.
This redshift effect is thus crucial in diminishing the particle production
rate under the small amplitude perturbation.
In most cases of cosmological application it is then important whether the
large amplitude perturbation may or may not occur,
since in this case the redshift effect plays a minor role.
It is shown that the large amplitude perturbation is realized at least
in one case examined:
when a large intrinsic coupling is
involved, which may or may not occur for the Polonyi field,
depending on supergravity models.
Using the formulas applicable under the large amplitude perturbation,
we examine how the produced energy density relative to the parent's
varies with cosmological evolution
and under what condition this ratio becomes of order unity,
in which case a catastrophic decay of WISF is expected.
It is suggested that in the Polonyi field case the catastrophic particle
production terminates the Polonyi field oscillation if  an intrinsic
coupling is larger than $\sim 10$.

%%%%%%%%%%%%%%%%%%%%%%%%%%%%%%%%%%%%%%%%%%

\vspace{2cm}

{\bf \lromn 2
\hspace{0.3cm}
Model in the Schr$\stackrel{..}{{\rm o}}$dinger picture}

\vspace{0.5cm}
In order to extract the most important features of the problem
and not to be confused
by minor details, we adopt a simple model of scalar boson field
$\varphi(\vec{x} \,, t) $
coupled to a classical oscillator $\xi(t) $ with the Hamiltonian,
\begin{eqnarray}
H =
\int\,d^{3}x\,[\:\frac{1}{2}\, \pi _{\varphi }^{2} + \frac{1}{2}\,
( \nabla \varphi )^{2} + \frac{1}{2}\,
m^{2}\varphi ^{2} - \frac{1}{2}\,
\mu  \xi(t) \varphi ^{2}\:] \,.
\end{eqnarray}
A slightly general form of time dependence of
the periodic oscillator $\xi (t)$ is taken here:
\begin{eqnarray}
\xi (t) = \xi_{+}\,e^{-\,2iz} + \xi_{-}\,e^{2iz} \,, \hspace{0.5cm}
z = \frac{E}{2}\,t \,,
\end{eqnarray}
with real constants $\xi _{\pm }$.
The oscillation frequency is given by $E$ and for a practical purpose
one can take an amplitude of
\( \:
\xi _{+} = \xi _{-} = \frac{\xi }{2} \,,
\: \)
so that
\( \:
\xi (t) = \xi\,\cos (Et) \,.
\: \)
The reason we take a more general form of
\( \:
\xi _{+} \neq \xi _{-}
\: \)
is to show that
none of our conclusion depends on the special cosine form, the result
resting solely with presence of the two terms of the form, $e^{\pm 2iz}$, in
any finite mixture.
On the other hand, a single term, either $e^{2iz}$ or $e^{-\,2iz}$,
does not work.
As mentioned in Introduction, we may envisage the situation
in which the oscillator
$\xi(t) $ is a homogeneous component of a genuine field variable
$\xi (\vec{x} \,, t)$.

In the rest of discussion, we take the
Schr$\stackrel{..}{{\rm o}}$dinger picture of the quantum Hilbert space.
The field variable $\varphi (\vec{x} \,, t)$ is decomposed into its
spatial Fourier components $q_{\vec{k} }( t)$,
and the wave functional of the state vector
is written as
\begin{eqnarray}
\Psi[\, \{\,q_{\vec{k}}\,\} \,, t\,] =
\prod_{\vec{k}}\,\psi_{\vec{k}} (q_{\vec{k}} \,, t) \,.
\end{eqnarray}
Due to the space translational invariance it is obvious that each Fourier
sector is linearly independent.
For definiteness we enclose the system within a large normalization box
of volume $V$ such that each mode is labeled by a discrete index $\vec{k}$.
The wave function $\psi_{\vec{k}}$
obeys the Schr$\stackrel{..}{{\rm o}}$dinger equation,
\begin{eqnarray}
i\,\frac{\;\partial \psi }{\partial t} = -\,\frac{1}{2}\,
\frac{\partial ^{2}\psi }{\partial q_{\omega }^{2}} + \frac{1}{2}\,
\omega ^{2}(t)\,q_{\omega }^{2}\,\psi \,, \label{schrodinger eq}
\end{eqnarray}
with
\begin{eqnarray*}
\omega ^{2}(t) = \omega ^{2} - \mu \xi (t) \,.
\end{eqnarray*}
To simplify our notation we denoted each Fourier component by the index
\( \:
\omega = \sqrt{\,\vec{k}^{2} + m^{2}\,}
\: \)
, and we shall omit its suffix $\omega $ whenever no confusion may occur.

The Schr$\stackrel{..}{{\rm o}}$dinger equation Eq.\ref{schrodinger eq}
is solved by a Gaussian ansatz,
\begin{eqnarray}
\psi (q_{\omega }\,, t) = \frac{1}{\sqrt{u}}\,\exp (\:\frac{i}{2}\,
\frac{\dot{u}}{u}\,q_{\omega }^{2} \:) \,,
\end{eqnarray}
where it is found that
\begin{eqnarray}
\frac{d^{2}\,u}{dt^{2}} + \omega^{2}(t)\,u= 0 \,.
\end{eqnarray}
The dot $\dot{}$ denotes the time derivative $\frac{d}{dt}$.
This equation is equivalent to the corresponding classical field equation
coupled to the oscillator $\xi $.
Namely, the quantum problem is converted to the classical problem by
the Gaussian ansatz.

By changing the variables like
\begin{eqnarray}
u(z) = u(t) \,, \hspace{0.5cm}
z = \frac{1}{2}\, E\,t \,, \hspace{0.5cm}
h = (\frac{2\omega }{E})^{2} \,, \hspace{0.5cm}
\theta_{\pm } = \frac{4\mu \xi _{\pm }}{E^{2}} \,,
\end{eqnarray}
we find that the classical equation is of a generalized Mathieu type,
\begin{eqnarray}
\frac{d^{2}\,u}{dz^{2}} + (\,h - \theta_{+}\,e^{-\,2iz} - \theta _{-}\,e^{2iz}
\,)\,u = 0 \,. \label{general mathieu eq}
\end{eqnarray}
Due to the reality of the constants $\theta _{\pm }$,
once a special solution  $f(z)$ to
the second order differential equation Eq.\ref{general mathieu eq} is found,
another linearly independent solution is obtained as $f^{*}(-z)$.
Furthermore, from general discussion \cite{mathieu eq}
of the Mathieu type equation
and as will also be explained in the next section,
the special solution is written as the form of
\( \:
e^{\lambda  z}\,P(z)
\: \)
where $P(z)$ is periodic:
\( \:
P(z + \pi ) = P(z) \,.
\: \)
The parameter $\lambda  $ is either real or purely imaginary, indicating
respectively unstable and bounded solution.
In mathematics these are generally called Floquet
solution, while in physics they are parametric resonance for $\lambda $ real
and Bloch wave for $\lambda $ purely imaginary
(with time replaced by a spatial coordinate).

Our main interest lies in evolution of the initial ground state,
\begin{eqnarray}
\psi (q \,, 0) = (\frac{\omega_{0} }{\pi })^{1/4}\,
e^{-\,\frac{1}{2}\, \omega_{0}\, q^{2}} \,, \hspace{0.5cm}
\omega _{0} = \sqrt{\,\omega ^{2} - \mu (\xi _{+} + \xi _{-})\,} \,,
\end{eqnarray}
hence we take as the initial condition to the classical differential equation,
\begin{eqnarray}
u(0) = (\frac{\omega_{0} }{\pi })^{-\,1/2} \,, \hspace{0.5cm}
\frac{\dot{u}}{u}\,(0) = i\omega_{0} \,.
\end{eqnarray}
With this initial condition
one may further achieve a simplification of the
wave function.
First, separating real and imaginary parts with
\( \:
u(z) = R(z) + iI(z) \,,
\: \)
one finds that
\begin{eqnarray}
&&
\frac{\dot{u}}{u} = \frac{E}{2}\,\frac{R' + iI'}{R + iI}
= \frac{E}{2}\,[\:\frac{R'R + I'I}{R^{2} + I^{2}} + i
\frac{I'R - R'I}{R^{2} + I^{2}}\:] \,, \nonumber
\end{eqnarray}
where the prime ' indicates the $z-$derivative
\( \:
\frac{d}{dz} \,.
\: \)
By noting the constancy of the Wronskian of the two solutions,
\( \:
R(z) \; {\rm and} \; I(z)
\: \),
\begin{eqnarray*}
R\dot{I} - \dot{R}I = (\frac{\omega_{0} }{\pi })^{-\,1}\,\omega_{0}  = \pi
\,,
\end{eqnarray*}
one obtains
\begin{eqnarray}
&&
\Re \frac{\dot{u}}{u} =
\frac{1}{2}\, \frac{d}{dt}\,\ln |u|^{2}
\,, \hspace{0.5cm}
\Im \frac{\dot{u}}{u} = \frac{\pi }{|u|^{2}} \,,
\nonumber \\
&& \hspace*{-2cm}
{\rm thus} \,, \hspace{0.5cm}
\psi (q\,, t) = \frac{1}{\sqrt{|u|}}\,\exp\,[\:-\,\frac{q^{2}}
{2|u|^{2}}\, \{\: \pi - \frac{i}{2}\,\frac{d}{dt}\,|u|^{2}\:\}
 \:] \,. \label{wave function}
\end{eqnarray}
We have ignored a $q-$independent phase factor that do not affect
physical observables.
This is a basic formula for our subsequent investigation.

The Gaussian wave function Eq.\ref{wave function}
of the evolved ground state is characterized by
the spread and a phase factor, the spread of the Gaussian peak being given
by the modulus $|u|$ that almost, but not quite completely
(rather $u$ itself faithfully), follows the classical motion
and the phase being given by the rate of time variation of this spread.
Thus in our formalism there exists an unambiguous
correspondance between the quantum wave function
and the classical motion.
Another nice feature of the formula Eq.\ref{wave function} is that
the normalization of the wave function
\begin{eqnarray*}
\int_{-\infty }^{\infty }\,dq\,|\,\psi (q\,, t)\,|^{2} = 1 \,,
\end{eqnarray*}
is automatically satisfied, ensuring
the unitary development of the quantum state.
Despite of the unitary evolution the ground state decays, and
the decay amplitude is given by
\begin{eqnarray}
&&
\langle \,0 \,|\,\psi (t) \,\rangle =
(\frac{\omega_{0} }{\pi })^{1/4}\,\int_{-\,\infty }^{\infty }\,dq\,
e^{-\,\omega_{0} \,q^{2}/2}\,\psi (q\,, t) \nonumber \\
&&
\hspace*{1cm}
=
(4\pi \omega_{0} )^{1/4}\,[\:\omega_{0} \,|u | + \frac{\pi }
{|u|} - i \frac{d|u|}{dt}\,\:]^{-\,1/2} \,.
\end{eqnarray}
Thus exponential growth of the classical variable $u$ in the instability
band of the parametric resonance implies directly the exponential decay
law of the initial quantum state.
The decayed state is accompanied by production of field quanta,
since the spread of the Gaussian peak implies excitation of higher harmonic
oscillator levels.
We shall have much more to say later on these issues
after we introduce the density matrix of the quantum system.

%%%%%%%%%%%%%%%%%%%%%%%%%%%%%%%%%%%%%%%%%%%%%%%%%%%%%%%%%%%%%

\vspace{2cm}

{\bf
\lromn 3 \hspace{0.3cm}
Density matrix of Gaussian model and its general properties}

\vspace{0.5cm}
Although the wave function completely describes a quantum state,
an equivalent description is possible in terms of the density matrix.
It is sometimes, for instance in cases when observable quantities are
needed, also useful to introduce the density matrix.
In our subsequent discussion on the precise nature of the quantum state
at late times, it is of crucial importance to use the density matrix formalism.
This is due to that the density matrix in the Gaussian model has some
simple properties, while preserving the fundamental relation of unitarity.

The entire density matrix of our quantum system
is decomposed into a product of submatrices $\rho _{\vec{k}}$ in Fourier
modes as
\begin{eqnarray}
\rho = \prod_{\vec{k}}\,\rho _{\vec{k}} \,, \hspace{0.5cm}
\rho _{\vec{k}}(q\,, q' \;; t) = \psi (q \,, t)\,\psi ^{*}(q' \,, t)
= \langle \,q \,|\,\psi (t) \,\rangle \,\langle \,\psi (t) \,|\,q' \,\rangle
\,.
\end{eqnarray}
Unless confusion occurs, we often omit the mode index $\omega $ and the
time variable $t$.
The wave function in our problem is of a Gaussian form,
\begin{eqnarray}
\psi (q \,, t) = (\frac{\Re D}{\pi })^{1/4}\,e^{-\,\frac{D}{2}\,q^{2}} \,,
\end{eqnarray}
with $D(t)$ a known time-dependent function in our analysis of this section.
All discussion that follows in this section is valid for an arbitrary
$D(t)$ provided that $\Re D > 0$.

It is convenient to introduce the Fock base of the unperturbed field
oscillators $q_{\omega } $,
\begin{eqnarray}
|\,n \,\rangle = \frac{1}{\sqrt{n!}}\,(\,a_{\omega }^{\dag}\,)^{n}\,
|\,0 \,\rangle \,,
\hspace{0.5cm}
\langle \,q_{\omega } \,|\,n \,\rangle = \frac{1}{\sqrt{n!}}\,
(\,\sqrt{\frac{\omega }{2}}\,q_{\omega }  + \frac{1}{\sqrt{2\omega }}\,
\frac{\partial }{\partial q_{\omega }} \,)^{n}\,\psi _{0}(q_{\omega }) \,,
\end{eqnarray}
with
\( \:
\psi _{0}(q_{\omega }) \propto e^{-\,\omega\, q_{\omega }^{2}/2}
\: \)
the ground state wave function.
The state vector in the Fock base is then expressed by
\begin{eqnarray}
\langle \,n \,|\,\psi(t)  \,\rangle
= (\frac{\omega\, \Re D}{\pi^{2} })^{1/4}\,\frac{1}{2^{n/2}\,\sqrt{n!}}\,
\int_{-\,\infty }^{\infty }\,dq\,H_{n}(\sqrt{\omega }q)\,e^{-\,\frac{1}{2}\,
(\omega + D)\,q^{2}} \,,
\end{eqnarray}
with $H_{n}(x)$ the Hermite polynomial of order $n$.
The state vector components
\( \:
\langle \:n \:|\:\psi(t)  \:\rangle
\: \)
trivially vanish for an odd integer $n$. In the even integer case,
using an identity on the Hermite polynomial \cite{bateman 2},
\begin{eqnarray}
\int_{-\,\infty }^{\infty }\,dy\,e^{-\,y^{2}}\,H_{2l}(xy) =
\sqrt{\pi }\,\frac{(2l)!}{l!}\,(x^{2} - 1)^{l} \,, \nonumber
\end{eqnarray}
we obtain a very useful relation,
\begin{eqnarray}
\int_{-\,\infty }^{\infty }\,dq\,H_{2l}(\sqrt{\omega }q)\,e^{-\,\frac{1}{2}\,
(\omega + D)\,q^{2}} =
\sqrt{\pi }\,\frac{(2l)!}{l!}\,\sqrt{\frac{2}{\omega + D}}\,
(\frac{\omega - D}{\omega + D})^{l} \,.
\end{eqnarray}
This formula greatly simplifies our subsequent discussion.

The density matrix in the Fock base is thus given by
\begin{eqnarray}
&&
\rho _{2l \,, 2m} = \frac{\sqrt{\:(2l)!\,(2m)! \:}}{2^{l+m-1}\,l!\,m!}\,
|\frac{\omega \,\Re D}{(\omega + D)^{2}}|^{1/2}\,
(\frac{\omega - D}{\omega + D})^{l}\,
(\frac{\omega - D^{*}}{\omega + D^{*}})^{m} \,, \\
&&
\rho _{n  m} = 0 \hspace{0.5cm} ({\rm for \; } n  \; {\rm or} \;
m \; {\rm odd}) \,.
\end{eqnarray}
It can be verified that this density matrix trivially
meets the unitarity relation,
\begin{eqnarray}
\sum_{n= 0}^{\infty }\, \rho _{nn} = {\rm tr}\;\rho = 1 \,, \nonumber
\end{eqnarray}
as should be so by construction.
The diagonal element of the sub-density  matrix is written in terms of
one parameter denoted by $K$,
\begin{eqnarray}
\rho _{2l \,, 2l} = \frac{(2l)!}{2^{2l}\,(l!)^{2}}\,
\sqrt{\frac{2K}{2+K}}\,(\,\frac{2 - K}{2 + K}\,)^{l} \,, \hspace{0.5cm}
K \equiv \frac{4\omega \,\Re D}{\omega^{2} + |D|^{2}} \,.
\label{diagonal denst matrix}
\end{eqnarray}

A few immediate conclusions follow from consideration given here.
The first and second moments of the density matrix give the average number
and the dispersion of produced particles, which is computed as
\begin{eqnarray}
&&
\langle \,N \,\rangle \equiv
\sum_{n= 0}^{\infty }\, n\,\rho _{nn} =
\frac{1}{4\omega \,\Re D}\,
|\:\omega - D \:|^{2}
=
\frac{2 - K}{2K}
\,, \\
&&
\langle \,\Delta N^{2} \, \rangle \equiv
\langle \,N^{2} \, \rangle - \langle \,N \, \rangle^{2} =
2\,|\:\frac{\omega + D}{\omega - D} \:|^{2}\,\langle \:N \: \rangle^{2} \,.
\end{eqnarray}
The further $D$ deviates from $\omega $, the more particles are produced.
The diagonal element of the density matrix Eq.\ref{diagonal denst matrix}
is expressed solely in terms of the average number of produced particles
\( \:
\langle \,N \,\rangle \,.
\: \)

The probability of the initial state being maintained is exactly given by
\begin{eqnarray}
|\,\langle \,0 \,|\,\psi (t) \,\rangle  \,|^{2} = \rho _{00} =
\sqrt{\frac{2K}{2+K}} \,.
\end{eqnarray}
A precise relation of the decay law of the initially prepared state and
the number of produced particles is then
\begin{eqnarray}
\rho _{00} = \frac{1}{\sqrt{\:1 +  \langle \,N \,\rangle \:}} \,.
\end{eqnarray}
Thus when the initial state decays according to
\( \:
e^{-\,\Gamma t} \,,
\: \)
the produced number of particles increases like
\( \:
e^{2\Gamma t} \,.
\: \)
This relation is exact irrespective of detailed form of the wave function,
but it must be kept in mind that this exact relation holds only for
a single frequency mode.
We shall later derive a formula applicable when the mode-sum is performed.

%%%%%%%%%%%%%%%%%%%%%%%%%%%%%%%%%%%%%%%%%%%%%%%%%%%%%%%%

\newpage
{\bf \lromn 4
\hspace{0.3cm}
Parametric resonance}

\vspace{0.5cm}
It is well known \cite{mathieu eq}
that the Mathieu type equation with periodic coefficients
exhibits instability in infinitely many band regions of the parameter space
of
\( \:
h = (\frac{2\omega }{E})^{2} \hspace{0.3cm} {\rm and} \hspace{0.3cm}
\theta_{\pm } = \frac{4\mu \xi _{\pm }}{E^{2}} \,.
\: \)
In the weak coupling, or the small amplitude limit of
\( \:
h \gg |\theta _{\pm }|
\: \)
the parametric resonance roughly occurs when the input frequency
$E$ is  closely tuned, within a resonance band,
to the characteristic frequency of the system,
$\omega $, by the relation:
\( \:
\omega = \frac{E}{2}\,n \,, \; {\rm namely} \,, \;
h = n^{2}
\: \)
with $n$ a positive integer.
The parametric resonance occurs however small the coupling
$\theta _{\pm }$ is.
Implication of this parametric resonance on the quantum ground state
is the main theme of this paper.
In this section we shall first review weak coupling analysis of the
classical parametric resonance,
explaining some details usually not written in the textbook
on Mathieu's equation.
We think that the material presented in the weak coupling limit either
includes new results of great use in our later analysis,
or is a new derivation of known
results in a much more simple way than usually presented.
In the second part of this section
we sketch analysis in the large amplitude or the
strong coupling limit.
For brevity we only present results to the first leading term in the
strong coupling limit.
In the last part of this section we combine analysis in the above two
regimes in order to identify strong coupling regions within a few lower
resonance bands, which becomes useful in our later application.
A precise identification of the adiabatic instability lines is difficult
to find, by analytic means, in a general ($ \theta \,, h $) region, but
for lower bands of $n=1 \,, 2 $ we can derive approximate analytic
formula by comparing the stability chart obtained by numerical methods
\cite{numerical mathieu} to our weak coupling result.

\vspace{0.5cm}
{\bf \lromn 4.A \hspace{0.2cm}
Small amplitude limit
}

\vspace{0.5cm}
A Floquet type of solution is  expanded as
\begin{eqnarray}
&& \hspace*{-2.5cm}
e^{\lambda  z}\,P(z)  =
\sum_{k = -\,\infty }^{\infty }\,c_{k}(z)\,e^{i(n+2k)z} \,,
\end{eqnarray}
with the periodicity understood:
\( \:
P(z+\pi ) = P(z) \,.
\: \)
The basic idea of our approach is that within the narrow band near
\( \:
h = n^{2}
\: \)
one may take $e^{\pm inz}$ as the zero-th order
approximate solution and assume
small time variation of the amplitude $c_{k}(z)$. It is thus legitimate to
keep time derivative terms only in the two modes,
\( \:
k = 0 \; {\rm and} -n \,,
\: \)
in the form of
\( \:
\dot{c}_{0 \,,-\,n} \sim \lambda  c_{0 \,,-\,n} \,,
\hspace{0.3cm} {\rm since} \;
|\lambda  | \ll 1
\,.
\: \)
We anticipate a systematic hierarchy of coefficients in the small
parameter $\sqrt{\theta _{+}\theta _{-}}$:
\begin{eqnarray*}
c_{k} = O[\,(\theta _{+}\theta _{-})^{{\rm Min.}\;(\,|k|/2 \,, |k+n|/2\,)}\,]
\,,
\end{eqnarray*}
whose validity can consistently be checked by solutions derived.
The growth rate factor $\lambda $ as a function of
\( \:
\epsilon = h - n^{2}
\: \)
is obtained as described in Appendix:
\begin{eqnarray}
\lambda  (\epsilon ) = \frac{1}{2n}\,\sqrt{\:
(\frac{\Delta \epsilon_{n} }{2})^{2}
- (\,\epsilon - \epsilon _{0} \,)^{2} \:}
\,.
\end{eqnarray}
Here we defined various quantities by
\begin{eqnarray}
\epsilon _{0} &=& \frac{1}{2}\, \frac{1}{n^{2}-1} \,\theta _{+}\theta _{-} \,,
\hspace{0.5cm} {\rm for} \hspace{0.2cm}  n \neq 1 \,, \nonumber \\
&=& -\,\frac{1}{8}\,\theta _{+}\theta _{-} \,, \hspace{0.5cm} {\rm for}
\hspace{0.2cm} n =1 \,, \\
\Delta \epsilon _{n} &=&
\frac{(\theta _{+}\theta _{-})^{n/2}}
{2^{2n-3}\,[\,(n-1)! \,]^{2}} \,. \label{band width}
\end{eqnarray}

The case of $n=0$ must be treated separately.
One finds that
\begin{eqnarray}
\lambda = \sqrt{\,-\,\frac{1}{4}\,\theta _{+}\theta _{-} - \epsilon \,}
\,, \hspace{0.5cm} {\rm with} \;
\epsilon < -\,\frac{1}{4}\,\theta _{+}\theta _{-} \,.
\label{n=0 line}
\end{eqnarray}
The instability region with a real $\lambda $ is characterized by infinitely
many narrow
bands with the width $\Delta \epsilon _{n}$ Eq.\ref{band width}.
However there is a gap
between the the $n=0$ instability line Eq.\ref{n=0 line}
and the first $n=1$ band:
this region is defined by
\begin{eqnarray}
-\,\frac{1}{4}\,\theta _{+}\theta _{-} < h < 1
- \sqrt{\,\theta _{+}\theta _{-}\,} - \frac{1}{8}\,\theta _{+}\theta _{-}
\,,
\end{eqnarray}
for a small $\theta _{+}\theta _{-}$.
In summary, the first few instability bands in the ($\theta \,, h$) plane
are, with
\( \:
\theta _{+} = \theta _{-} = \theta \,,
\: \)
\begin{eqnarray}
h < -\,\frac{\theta ^{2}}{4} \,, \hspace{0.3cm}
1 - \theta - \frac{\theta ^{2}}{8} < h < 1 + \theta - \frac{\theta ^{2}}{8}
\,, \hspace{0.3cm}
4 - \frac{\theta ^{2}}{12} < h < 4 + \frac{5}{12}\,\theta ^{2} \,.
\label{boundary line}
\end{eqnarray}
These formulas are accurate for small $\theta $ values.
At $\theta $ as large as $3$ the accuracy gets worse, but typically
down to $\sim 25\%$, which is still useful in many cases.

In terms of the original oscillator variables the resonance band lies
around
\( \:
\omega = \frac{E}{2}\,n
\: \)
with the band width given by
\begin{eqnarray}
\Delta \omega _{n} = \frac{(\mu \xi / E^{2})^{n}}
{2^{n-1}\,n\,[\,(n-1)!\,]^{2}}\,E \,,
\end{eqnarray}
taking
\( \:
\xi _{+} = \xi _{-} = \frac{1}{2}\, \xi
\: \)
for simplicity.
The gap of stability region around the origin in the
\( \:
(\,\sqrt{\mu \xi } \,, \omega \,)
\: \)
plane
is specified by the three lines,
\begin{eqnarray*}
\omega = 0 \,, \hspace{0.3cm} \mu \xi = 0 \,, \hspace{0.3cm}
\omega = \frac{m_{\xi }}{2}\,[\,1 - 2\,\frac{\mu \xi }{m_{\xi }^{2}}
- \frac{1}{2}\, (\frac{\mu \xi }{m_{\xi }^{2}})\,]^{1/2} \,.
\end{eqnarray*}
We call this region bounded by the three lines stability triangle.

It is also possible to compute expansion coefficients of the solution
$u(z)$ in terms of a single overall factor $c_{0}$.
We quote the leading result,
\begin{eqnarray}
u
\sim
c_{0}\,e^{\lambda z + inz} + c_{-\,n}\,e^{\lambda z - inz}
=
\frac{\theta _{-}^{n/2}}{\sqrt{\:\theta _{+}^{n} + \theta _{-}^{n}\:}}\,
e^{\lambda  z + i(nz - \eta )} +
\frac{\theta _{+}^{n/2}}{\sqrt{\:\theta _{+}^{n} + \theta _{-}^{n}\:}}\,
e^{\lambda  z - i(nz - \eta )} \,,
\end{eqnarray}
with
\begin{eqnarray}
\tan \eta = \sqrt{\frac{\epsilon _{+}^{(n)} - \epsilon }
{\epsilon - \epsilon _{-}^{(n)}}} \,, \hspace{0.5cm}
\epsilon _{\pm }^{(n)} = \epsilon _{0} \pm \frac{\Delta \epsilon_{n} }{2} \,.
\end{eqnarray}

\vspace{0.5cm}
{\bf \lromn 6.B \hspace{0.2cm}
Large amplitude limit}

\vspace{0.5cm}
Having presented the necessary machinery
in the weak coupling or the small amplitude limit,
we proceed to the strong coupling or the large
amplitude case. Here our discussion is limited to the leading approximation.
Consider the Mathieu equation in a slightly different, but physically
equivalent fashion:
\begin{eqnarray}
u'' + \tilde{\omega }^{2}(z)\,u = 0 \,, \hspace{0.5cm}
\tilde{\omega }^{2}(z) \equiv h + 2\theta \cos (2z) \,,
\end{eqnarray}
with the initial condition,
\begin{eqnarray}
u(0) = (\frac{\tilde{\omega }(0)E}{2\pi })^{-\,1/2} \,, \hspace{0.5cm}
\frac{u'}{u}(0) = i\tilde{\omega }(0) \,. \nonumber
\end{eqnarray}
We took both
\( \:
\theta > 0 \,, \; h > 0 \,.
\: \)
Assumption of the initial ground state of harmonic oscillators means that
\( \:
\tilde{\omega }(0) = \sqrt{\,h + 2\theta \,}
\: \)
is real and positive.
For simplicity we take
\( \:
\xi _{+} = \xi _{-}
\: \)
such that
\begin{eqnarray*}
\theta _{+} = \theta _{-} = \frac{2\mu \xi }{E^{2}} \,.
\end{eqnarray*}

The condition of the strong coupling or the large amplitude is
\( \:
2\theta \geq  h
\: \)
and we anticipate that this is realized by the adiabatic condition,
\begin{eqnarray}
|\frac{u'}{u}| > |\frac{\omega '}{\omega }| \,. \label{adiabatic condition}
\end{eqnarray}
Neglecting the right-hand side of
\begin{eqnarray}
\frac{d}{dz}\,[\,\frac{1}{2}\, (u')^{2} + \frac{1}{2}\,\tilde{\omega }^{2}
\,u^{2} \,] = \tilde{\omega} ' \tilde{\omega}\, u^{2} \,,
\end{eqnarray}
that follows from the classical equation of motion,
we find, by considering the initial condition,
\begin{eqnarray}
u(z) = (\frac{E\sqrt{h+2\theta }}{2\pi })^{-\,1/2}\,
\exp [\,i\int_{0}^{z}\,dz\,\sqrt{\,h + 2\theta - 4\theta \sin ^{2}z\,}\,] \,.
\label{strong amp}
\end{eqnarray}
The adiabatic condition Eq.\ref{adiabatic condition}
is obeyed for $\theta \gg h $ as expected, except near the turning point
of $\omega(z) = 0$.

The exponent factor in the formula Eq.\ref{strong amp}
is an elliptic integral $E(z \,, k)$ of the second kind,
\begin{eqnarray}
i\sqrt{h+2\theta }\,\int_{0}^{z}\,dz\,\sqrt{\,1 - k^{2}\,\sin ^{2}z\,}
= i\sqrt{h+2\theta }\,E(z \,, k) \,.
\end{eqnarray}
with
\begin{eqnarray}
k^{2} \equiv \frac{4\theta }{2\theta + h} \,,
\end{eqnarray}
which approaches $2$ as
\( \:
\theta \:\rightarrow  \: \infty \,.
\: \)
The behavior of $E(z \,, k)$ as a function of $z$ for $k > 1$
is best studied by the complex integral in the cut $w = \sin z$ plane:
\begin{eqnarray}
E(z \,, k) = \int_{0}^{\sin z}\,dw\,\sqrt{\frac{1-k^{2}w^{2}}{1-w^{2}}}\,.
\nonumber
\end{eqnarray}
Near the turning point
\( \:
w = \pm \frac{1}{k}
\: \)
one must analytically continue $w$ into the complex
upper half-plane so that one picks up the growing component which always
dominates over the decaying component.
Thus each time when $z$ goes beyond the branch point singurality
$\pm \sin ^{-\,1} \frac{1}{k}$,
the exponent gets enhanced, and the total increment in passing through
one period of $\Delta z = 2\pi $ is
\begin{eqnarray}
i\Delta E(2\pi \,, k)
= 4\,\int_{1/k}^{1}\,dw\,\sqrt{\frac{k^{2}w^{2}-1}{1-w^{2}}} \,.
\end{eqnarray}
In the limit of
\( \:
\theta \:\rightarrow  \: \infty \,,
\: \)
namely
\( \:
k \:\rightarrow  \: \sqrt{2} \,,
\: \)
this increment is numerically $\sim 2.396$.
In the large time limit of $z \gg 2\pi $ the growth factor accumulates,
and one finds that
\begin{eqnarray}
|u|^{2} \simeq \sqrt{\frac{2}{\theta }}\,\frac{\pi }{E}\,e^{1.2\sqrt{2\theta }
\,Et}
\,. \label{late time adiabatic}
\end{eqnarray}
When this formula is extended to an extreme, one may use it for the
whole range of
\( \:
0 < \omega < \sqrt{\mu \xi } \,,
\: \)
although one must be careful of the stability triangle mentioned above.
Furthermore, the important exponent factor has a variation at a finite
$h/(2\theta )$. As a function of this variable the constant
coefficient $1.2$ in Eq.\ref{late time adiabatic} is modified to
\begin{eqnarray}
J(\delta ) =
4\,(\,1 - \frac{2}{\pi }\,{\rm arcsin}\,\sqrt{\frac{1+\delta }{2}}\,)
\,\int_{\sqrt{\frac{1+\delta }{2}}}^{1}\,dw\,\sqrt{\,\frac{2w^{2}-1-\delta }
{1-w^{2}}\,} \,, \hspace{0.5cm}
\delta = \frac{h}{2\theta } \,.
\label{exponent rate}
\end{eqnarray}
The average value of this quantity for the range of
$ \delta = 0 - 1$ is $\sim 0.327$, a factor $\sim 0.3$ reduction from
$1.2$,
when averaged with the phase space weight.
We should consider this reduction for a better estimate.

\vspace{0.5cm}
{\bf \lromn 4.C \hspace{0.2cm}
Strong coupling region within lower bands}

\vspace{0.5cm}
The large amplitude or the strong coupling formula above has been derived
in the adiabatic approximation, with the condition
\( \:
h < 2\theta \,.
\: \)
But the actual region of applicable adiabatic region cannot be this
simple, especially due to existence of infinitely many bands.
{}From an inspection of the stability chart obtained by numerical methods
\cite{numerical mathieu}
one can however gain a confidence of deriving approximate analytical expression
of the instability region in which the adiabatic formula may be
applied, at least for a few lower bands of
\( \:
n = 1 \,, 2 \,.
\: \)

We proceed as follows.
First, we define the region bounded by the weak coupling band lines and
the adiabatic line of $h=2\theta $;
\begin{eqnarray}
n^{2} + \epsilon _{-}^{(n)} < h < {\rm Min.}\;(n^{2} + \epsilon _{+}^{(n)}\,,
2\theta ) \,, \hspace{0.5cm}
\epsilon _{\pm }^{(n)} \sim  \frac{1}{2}\, \frac{\theta ^{2}}{n^{2} - 1}
\pm  \frac{\Delta \epsilon _{n}}{2}\,, \hspace{0.2cm} (n \neq 1) \,.
\end{eqnarray}
Suppose then that $\theta $ gradually increases from $0$.
For $\theta < 0.33$ the gap of the stability triangle exists and
there is no adiabatic region. As $\theta $ increases above $0.33$,
the adiabatic $h = 2\theta $ line may cross the band boundary.
Each time the adiabatic $h=2\theta $ line crosses the lower weak
coupling line $h=n^{2} + \epsilon _{-}^{(n)}$, a new adiabatic region
appears. By using the weak coupling formula Eq.\ref{boundary line}
of the boundary line of the band,
we obtain approximate expressions of the strong coupling regions
within a few lower bands.
These regions are classified according to the value of $\theta $;
\begin{eqnarray}
&&
(1) \hspace{0.5cm}
2\,(\sqrt{38} -6) \;(\sim 0.33) < \theta < 2\,(\sqrt{6} - 2)\; (\sim 0.90) :
\nonumber \\
&& \hspace*{2cm}
1 - \theta - \frac{\theta ^{2}}{8} < h < 2\theta\,,  \hspace{0.2cm}
{\rm or} \nonumber \\
&& \hspace*{1cm}
\frac{m_{\xi }}{2}\,[\, 1 - 2\frac{\mu \xi }{m_{\xi }^{2}}
- \frac{1}{2}\, (\frac{\mu \xi }{m_{\xi }^{2}})^{2}\,]^{1/2}
< \omega < \sqrt{\mu \xi }
: \hspace{0.5cm} n = 1 \;{\rm band}  \; ({\rm \lromn 1}) \,, \\
&&
(2) \hspace{0.5cm}
0.90 < \theta < 3 : \nonumber \\
&& \hspace*{2cm}
0 < h < 1 + \theta - \frac{\theta ^{2}}{8}\,,  \hspace{0.2cm}
{\rm or} \nonumber \\
&& \hspace*{1cm}
0 < \omega <
\frac{m_{\xi }}{2}\,[\, 1 + 2\frac{\mu \xi }{m_{\xi }^{2}}
- \frac{1}{2}\, (\frac{\mu \xi }{m_{\xi }^{2}})^{2}\,]^{1/2}
: \hspace{0.5cm} n = 1 \;{\rm band} \; ({\rm \lromn 2}) \,, \\
&& {\rm for} \; \theta > 2\,(4\sqrt{3} - 6) \;(\sim 1.86) \,, \hspace{0.5cm}
4 - \frac{\theta ^{2}}{12} < h < 2\theta \,, \hspace{0.2cm}
{\rm or} \nonumber \\
&& \hspace*{1cm}
\frac{m_{\xi }}{2}\,[\,4 - \frac{1}{3}\,(\frac{\mu \xi }{m_{\xi }^{2}})\,]
^{1/2} < \omega < \sqrt{\mu \xi }
: \hspace{0.5cm} n = 2 \;{\rm band} \; ({\rm \lromn 3}) \,.
\end{eqnarray}
We call these three regions, Region \lromn 1, Region \lromn 2, Region
\lromn 3 for later application.
There is no simple analytic formula for
\( \:
\theta > 3 \,.
\: \)
But for a $\theta $ larger than $O[10]$ most of the region given by
\( \:
0 < h < 2\theta
\: \)
belongs to the instability region.

\vspace{2cm}

{\bf \lromn 5 \hspace{0.3cm}
Limiting behavior of evolved ground state
}

\vspace{0.5cm}
In this section we combine our previous formalism and mathematical analysis
in order to determine the final stage of initially prepared ground state
when the system is coupled to a periodic perturbation of frequency
 $E$. The sub-density matrix
\( \:
\rho _{\omega }
\: \)
with
\( \:
\omega = \sqrt{\,\vec{k}^{2} + m^{2}\,}
\: \)
of the quantum system is  a bounded function of time for most of the modes,
except within the infinitely many resonance bands, each joined to
\( \:
\omega = \frac{E}{2}\,n  \; (n = 1\,, 2 \,, 3 \,, \cdots )
\: \)
at the weak coupling limit of $\theta _{+}\theta _{-} = 0$.
Thus the limiting form of the density matrix as
\( \:
t \:\rightarrow  \: \infty
\: \)
is dominated by contribution  from resonance frequencies.
We therefore concentrate on combined effects from infinitely many bands,
neglecting contribution outside the instability region.

Recall, prior to any detailed discussion,
that once the classical field oscillator amplitude $u$ is given,
non-vanishing sub-density matrix element $\rho _{nm}$ in the Fock base of
frequency $\omega $ is obtained as
\begin{eqnarray}
&& \hspace*{-1cm}
\rho _{2l\,, 2m} =
\frac{\sqrt{\:(2l)!\,(2m)! \:}}{2^{l+m-1}\,l!\,m!}\,f_{lm} \,, \hspace{0.5cm}
f_{lm} =
|\frac{\omega\, \Re D}{(\omega + D)^{2}}|^{1/2}\,
(\frac{\omega - D}{\omega + D})^{l}
\,(\frac{\omega - D^{*}}{\omega + D^{*}})^{m} \,, \\
&&
D = \frac{\pi }{|u|^{2}} + \frac{i}{2}\,\frac{d}{dt}\,
\ln \frac{1}{|u|^{2}} \,.
\end{eqnarray}

\newpage
{\bf \lromn 5.A \hspace{0.2cm}
Small amplitude case}

\vspace{0.5cm}
The leading small amplitude
\( \:
(\,\theta  \ll  h \,)
\: \)
behavior is computed from the previous
general formula by puting in the initial condition.
For simplicity we present results taking
\( \:
\theta _{+} = \theta _{-} = \frac{2\mu \xi }{E^{2}} \,.
\: \)
For the $n-$th band it reads:
\begin{eqnarray}
&&
u(z) \simeq i\,(\frac{\omega_{0} }{\pi })^{-\,1/2}\,
\nonumber \\
&&
\frac{1}{\sin (2\eta )}\,
[\:e^{inz}\,\{ \cos (2\eta )\,
\sinh (\lambda  z) - i\sin (2\eta )\,\cosh (\lambda  z) \}
+ e^{-\,inz}\,\sinh (\lambda  z) \:] \,, \\
&& \hspace*{-2cm}
\lambda  =
\frac{\sqrt{\,(\epsilon _{+} - \epsilon )(\epsilon - \epsilon _{-})\,}}
{2n} \,,  \hspace{0.5cm}
\epsilon = (\frac{2\omega }{E})^{2} - n^{2}
\,, \hspace{0.5cm}
\sin (2\eta ) = 2\,
\frac{\sqrt{\,(\epsilon _{+} - \epsilon )(\epsilon - \epsilon _{-})\,}}
{\Delta \epsilon _{n}}
\,, \\
&&
\Delta \epsilon _{n} =
\frac{(\mu  \xi /E^{2})^{n}}
{2^{n-3}\,[\,(n-1)!\,]^{2}} \,, \hspace{0.5cm}
\epsilon _{\pm } = \frac{1}{2}\, \frac{\theta ^{2}}{n^{2} - 1} \pm
\frac{\Delta \epsilon _{n}}{2} \hspace{0.2cm} (n \neq 1) \,,
\end{eqnarray}
for
\( \:
\epsilon _{-} < \epsilon < \epsilon _{+}
\: \)
with
\( \:
z = \frac{E}{2}\,t \,.
\: \)
Although it appears subleading, terms of order
\( \:
e^{-\,\lambda  z} = e^{-\,\lambda  Et/2}
\: \)
must be retained to get correct, non-divergent contribution near the threshold
of the band,
\( \:
\epsilon \sim \epsilon _{\pm } \,.
\: \)

As emphasized in the previous section,  the density matrix element in each
mode is governed by the two numbers,
\( \:
K \sim 1/\langle \,N\,\rangle
\: \)
and a phase $\delta $ with
\( \:
\tan \delta = \Im D/\omega \,.
\: \)
Time dependence of this factor $K^{-\,1} \sim \langle N \rangle$ is
necessarily complicated, because there are two different time scales
involved, one scale very short of order $1/E$ and the other scale very long
of order $1/(\lambda  E)$. Note that
\( \:
\lambda  = O[\:\Delta \epsilon  \:] \ll 1 \,.
\: \)
A useful quantity is time average over the short
time scale, for instance by averaging over a quasi-period of
\( \:
\Delta t = \frac{2\Delta z}{E} = \frac{2\pi }{E} \,.
\: \)
Although exact time average of any density matrix element is in general
difficult to compute in a closed form,
it is relatively easy to get the average of the first moment,
\begin{eqnarray}
\overline{\,\langle N \rangle \,} = \overline{K^{-\,1}}
\simeq
\frac{\sinh ^{2}(\lambda  z)}{\sin ^{2}(2\eta )} \,.
\end{eqnarray}
This was derived by using the technique of complex integration like
\begin{eqnarray}
\frac{1}{\pi }\,\int_{z}^{z+\pi }\,dz\,f(\,e^{2i(nz - \eta )}\,) =
\frac{1}{2\pi i}\,\oint_{|w| = 1}\,\frac{dw}{w}\,f(w) \,. \nonumber
\end{eqnarray}

The number of produced particles thus computed has a correct finite behavior
near the ends of the resonance band at
$\epsilon = \epsilon _{\pm }$, since the dangerous factor
\( \:
(\epsilon _{+} - \epsilon )\,(\epsilon - \epsilon _{-})
\: \)
in
\( \:
\sinh ^{2}(\lambda z)
\: \)
and
\( \:
\sin ^{2}(2\eta )
\: \)
cancel in the time average
\( \:
\overline{\,\langle N \rangle \,} \,.
\: \)
It is thus possible to get a finite result by keeping small
\( \:
O[\,e^{-\,\lambda  Et/2}\,]
\: \)
terms consistently.
One indeed gets a well-defined number of produced particles
by summing over all unstable modes within the $n-$th band,
\begin{eqnarray}
&&
\sum_{\omega \sim nE/2}\,\overline{\,\langle N_{\omega } \rangle \,}
\simeq  \frac{E^{3}\,V}{32\pi ^{2}}\,
\sqrt{\:n^{2} - \frac{4m^{2}}{E^{2}}\:}\,\int_{\epsilon _{-}}^{\epsilon _{+}}
\,d\epsilon \,\overline{\,\langle N_{\omega } \rangle \,}
\nonumber \\
&& \hspace*{-1cm}
\sim
\frac{E^{3}\,V}{2^{8}\,\sqrt{2}\,\pi ^{3/2}}\,
\sqrt{\:n\,(\,n^{2} - \frac{4m^{2}}{E^{2}}\,) \:}
\,\frac{\sqrt{\,\Delta \epsilon_{n}\,}
 \,\exp [\: \Delta \epsilon_{n} \,Et/(4n)\:]\,}
{\sqrt{\,Et\,}} \,, \label{intensity rule}
\end{eqnarray}
where we explicitly wrote $n-$dependence.
An approximate estimate of integral,
\begin{eqnarray*}
\int_{\epsilon _{-}}^{\epsilon _{+}}\,d\epsilon \,
\frac{\sinh ^{2}(\lambda  z)}{\sin ^{2}(2\eta )} =
(\Delta \epsilon )^{2}\,\int_{0}^{\Delta \epsilon /2}\,dy\,
\frac{\sinh ^{2}(\frac{yz}{2n})}{y\,\sqrt{\,(\Delta \epsilon )^{2} - 4y^{2}\,}}
\sim
\frac{1}{4}\,\sqrt{\frac{\pi n\Delta \epsilon }{z}}\,e^{\Delta \epsilon z/(2n)}
\,,
\end{eqnarray*}
valid for
\( \:
\Delta \epsilon z/(2n) \gg 1
\: \)
was used here.
We define the growth rate of produced particles by
\begin{eqnarray}
\gamma _{n} = \frac{E}{4n}\,\Delta \epsilon _{n} =
\frac{(\mu \xi /E^{2})^{n}}{2^{n-1}\,n\,[(n-1)!]^{2}}\,E \,.
\end{eqnarray}
In the special case of $n=1$
\begin{eqnarray}
&&
\sum_{\omega \sim E/2}\,\overline{\,\langle N_{\omega } \rangle \,}
\sim
\frac{E^{3}\,V}{2^{8}\,\pi ^{3/2}}\,
\sqrt{\:1 - \frac{4m^{2}}{E^{2}} \:}
\,\sqrt{\,\frac{\theta }{Et}\,}\,e^{\gamma _{1}\,t} \,,
\hspace{0.5cm}
\gamma _{1} = \frac{\theta E}{2} = \frac{\mu \xi }{E^{2}} \,.
\end{eqnarray}

Time dependence of the spectrum and the intensity of produced particles
is of crucial importance in any application of the resonance production.
The spectrum has an unmistakable feature of equal spacing,
\( \:
\omega = \frac{E}{2}\,n \,,
\: \)
and of the band width,
\begin{eqnarray}
\Delta \omega _{n}
=
\frac{2\,(\mu  \xi/E^{2})^{n}}{n\,[\,(n-1)!\,]^{2}}\,E \,. \nonumber
\end{eqnarray}
Moreover, we explicitly gave the intensity rule at late times in Eq.
\ref{intensity rule}.
If this phenomenon should be verifiable at laboratories, it has
many experimental consequences that can be checked with detailed prediction.
The total number of produced particles at any finite time, summed over
all band index $n$, is finite, but becomes exponentially large at late
times.

The dominant contribution out of infinitely many bands is from
the fundamental mode of $n=1$, the case of pair production of equal energy
$\frac{E}{2}$. The dominance of the fundamental mode is true
except in the large amplitude case of
\( \:
 \theta = 2\mu \xi/E^{2} = O[1] \,,
\: \)
for which many pairs of higher energy may readily be created.
We shall have more to say on the large amplitude case
in the second part of this section.
Production of high energy particles may superficially appear to violate
the energy conservation,
$\omega > \frac{E}{2}$.
The apparent violation is allowed due to that the external oscillation
makes the concept of energy conservation meaningless at a formal level.
However minor it may appear in the small amplitude case,
it is important to keep in mind that  due to their peculiar nature
high energy particles may play important roles.

The decay law of the initial vacuum is derived from
\begin{eqnarray} \hspace*{-1cm}
\prod_{\omega }\,\rho _{00}(\omega ) \simeq
 \prod_{\omega }\,\sqrt{K_{\omega }}
\simeq
\exp [\:-\,\frac{1}{2}\, \sum_{\omega }\,\ln \langle N _{\omega }\rangle \:]
\:\propto  \: \exp [\:-\,\frac{1}{2}\, \sum_{\omega }\,
\lambda  _{\omega }\,E\,t \:]
\equiv e^{-\,\Gamma Vt} \,, \label{vacuum decay}
\end{eqnarray}
which yields
\begin{eqnarray} \hspace*{-1cm}
\Gamma = \sum_{n}\,\Gamma _{n} \,, \hspace{0.5cm}
\Gamma _{n} = \frac{E}{2V}\,\sum_{\omega \sim En/2}\,\lambda  (\omega ) =
\frac{E^{4}}{16\pi n}\,\sqrt{n^{2} - \frac{4m^{2}}{E^{2}}}\,
\frac{(\mu \xi /E^{2} )^{2n}}{2^{2n}\,[(n-1)!]^{4}}
\,. \label{decay rate}
\end{eqnarray}
$\Gamma $ is the total decay rate of the initial vacuum per unit
volume.
By taking the dominant $n=1$ mode alone, one gets
\begin{eqnarray}
\Gamma \sim 5.0 \times 10^{-3}\,(\mu \xi )^{2} \,,
\end{eqnarray}
for $m \ll E/2$.
The fact that the probability of the vacuum persistence Eq.\ref{vacuum decay}
is suppressed
by the volume factor $V$ in the exponent is related to the homogeneity
of space: the decay may occur anywhere with equal probability.
The decay rate Eq.\ref{decay rate}
may be compared with the Born rate of the usual one particle decay
of mass $M$ that conserves energy,
\begin{eqnarray}
\gamma = \frac{\mu ^{2}}{32\pi M}\,\sqrt{1 - \frac{4m^{2}}{M^{2}}} \,.
\nonumber
\end{eqnarray}
This formula can also be derived from the above partial rate
\begin{eqnarray*}
\Gamma _{1}\,V = \frac{V\,\mu ^{2}\,\xi^{2}}{64\pi }\,
\sqrt{\,1 - \frac{4m^{2}}{E^{2}}\,} \,,
\end{eqnarray*}
by taking with $ E = M $ the amplitude of
\( \:
\xi  = \sqrt{\frac{2}{MV}} \,,
\: \)
which follows by equating the energy density of $\xi -$oscillation
\( \:
\frac{1}{2}\, M^{2}\,\xi ^{2}
\: \)
to one particle mass density within the volume $V$,
\( \:
M/V \,.
\: \)
Difference here from this conventional Born formula lies in that
the amplitude of oscillation $\xi$ may become
much larger than that of one particle decay; indeed, the rate is enhanced
by $V\cdot M\xi^{2}/2$ which may become a huge factor in
cosmological application.

\vspace{0.5cm}
{\bf \lromn 5.B \hspace{0.2cm}
Large amplitude case}

\vspace{0.5cm}
We quote very briefly the main result.
For simplicity we take
\( \:
\xi _{+} =\xi _{-} = \frac{1}{2}\, \xi
\: \)
as before.
First, the produced number of particles for each mode
is given, using numerical values of the integral, by
\begin{eqnarray}
\langle \,N\,\rangle
\simeq 0.61\,\exp [\,2\,J(\frac{\omega ^{2}}{\mu \xi })\,\sqrt{\mu \xi }\,t\,]
 \,, \hspace{0.5cm}
2J(0) \sim 2.4
\,.
\end{eqnarray}
As explained in Section \lromn 4.B, the rate factor has a mode frequency
dependence given by $J(\omega ^{2}/\mu \xi )$ with $J(\delta )$ defined
by Eq.\ref{exponent rate}.
The mode sum is performed with the instability constraint, which
restricts the momentum range. If the whole phase space is available with
the condition $h<2\theta $ alone, then
\( \:
\vec{k}^{2} < \mu \xi - m^{2} \,,
\: \)
but the adiabatic regime is restricted in more complicated ways
as described in the previous section.
Here we merely give an order of magnitude estimate by taking the full
phase space and ignoring the $\omega-$dependence of the exponent.
We postpone discussion of the actual
reduction factor of the phase space to when application is made.
Thus, we find for production of  massless particles with
\( \:
m \ll \sqrt{\mu \xi} \,,
\: \)
\begin{eqnarray}
&&
\sum_{\omega }\,\frac{\langle \,N\,\rangle }{V} \simeq
1.0\times 10^{-2}\,(\mu \xi )^{3/2}\,e^{2.4\,\sqrt{\mu \xi }\,t} \,,
\\
&& \hspace*{-2cm}
{\rm energy \; density} \hspace{0.5cm}  \rho \simeq
0.77 \times 10^{-2}\,(\mu \xi )^{2}\,e^{2.4\,\sqrt{\mu \xi }\,t} \,.
\end{eqnarray}
{}From
\begin{eqnarray}
\rho _{00} \simeq \prod_{\omega }\,\frac{1}{\sqrt{\,\langle \,
N_{\omega }\,\rangle \,}} \;\propto\; \exp [\,-\,\frac{Et}{2}\,\sum_{\omega }\,
\lambda _{\omega }\,] \,, \nonumber
\end{eqnarray}
the decay law of vacuum is
\begin{eqnarray}
\rho _{00} \:\propto  \:e^{-\,\Gamma Vt} \,, \hspace{0.5cm}
{\rm with} \hspace{0.3cm}
\Gamma \sim \frac{(\mu \xi )^{2}}{6\pi ^{2}}\,
\langle \,\frac{\lambda _{\omega }}{\sqrt{2\theta }}\,\rangle \simeq
2.4 \times 10^{-2}\,(\mu \xi )^{2} \,,
\end{eqnarray}
again taking the full phase space.

%%%%%%%%%%%%%%%%%%%%%%%%%%%%%%%%%%%%%%%%%%%%%%%%%%55

\newpage

{\bf \lromn 6 \hspace*{0.2cm}
Application: fate of cosmological WISF
}

\vspace{0.5cm}
Modern unified theories predict
many spinless bose fields that couple to ordinary matter extremely weakly.
In cosmology based on these unified theories, there appears
a grave potential problem.
The condensate of these fields at the early universe
may dominate the mass density of the universe after some epoch of
oscillation, yet
they may or may not decay due to their weak coupling.
In either case they are likely to cause a disaster, because they either
produce too much entropy when they decay at too late an epoch,
destroying the standard nucleo-synthesis or the
baryo-genesis scenario, or their mass density overcloses the present
universe, which cannot be true.
We shall examine how the new mechanism of resonant particle production
discussed in previous sections affects this problem.
In particular, we pay special attention to the case of the invisible axion
\cite{axioncosmo},
the Polonyi \cite{polonyi}, and the modular fields,
which may be of immediate interest.

Weakly interacting spinless fields (WISF),
which we generically denote by $\xi $,
undergo damped oscillation at the early universe,
obeying the evolution equation,
\begin{eqnarray}
\stackrel{..}{\xi } + 3\frac{\dot{a}}{a}\,\dot{\xi } + m^{2}_{\xi }\,\xi =0
\,,
\end{eqnarray}
with $a(t)$ the cosmological scale factor in the flat Robertson-Walker
metric of
\( \:
ds^{2} = dt^{2} - a^{2}(t)\,d\vec{x}^{2} \,.
\: \)
For
\( \:
m_{\xi }\,t \gg   1 \,,
\: \)
the amplitude of $\xi $ field behaves as
\begin{eqnarray}
\xi (t) = \xi _{0}\,[\,\frac{a(t_{0})}{a(t)}\,]^{3/2}\,
\cos [\,m_{\xi }\,(t-t_{0})\,]
= \xi _{0}\,(\,\frac{T}{T_{0}}\,)^{3/2}\,\cos [\,m_{\xi }\,(t-t_{0})\,]
\,,
\end{eqnarray}
with
\( \:
t_{0} = 1/m_{\xi } \,,
\: \)
and $T_{0}$ the temperature scale of the onset of damped oscillation.
The initial temperature $T_{0}$ is determined by equating the Hubble time scale
to the inverse mass of $\xi $ field, $1/m_{\xi }$,
\begin{eqnarray}
T_{0} = (\frac{3}{8\pi d})^{1/4}\,
\sqrt{\,m_{\xi }\,m_{{\rm pl}}\,} \,, \hspace{0.5cm}
d = \frac{\pi ^{2}}{30}\,N \,,
\end{eqnarray}
where $N$ is an effective number of massless species contributing to the energy
density at the temperature $T_{0}$.

We consider a classs of WISF's that include the axion,
the Polonyi, and the modular fields. Their properties relevant to our
discussion are summarized by
a few quantities, the mass $m_{\xi }$, the initial amplitude of oscillation
$\xi _{0}$, and their generic Yukawa-type ($\frac{1}{2}\, \xi \varphi ^{2}$)
coupling denoted by $\mu $, to a typical
ordinary bose matter $\varphi $ such as
the photon, the Higgs, and the scaler fermion fields.
To the Hubble time scale the damping of oscillation amplitude is small,
and one may assume exactly periodic oscillation.
Mode functions of the bose field $\varphi $ obey
\begin{eqnarray}
\stackrel{..}{q}_{\omega } + [\,\omega ^{2} - \mu \xi (t) \,]\,q_{\omega } = 0
\,, \hspace{0.5cm}
\omega ^{2} = \vec{k}^{2} + m^{2}
\,,
\end{eqnarray}
during one Hubble time.
Rewriting this equation in the standard form of Mathieu type
gives the parameter set,
\begin{eqnarray}
E = m_{\xi } \,, \hspace{0.5cm}
\theta = \frac{2\mu \xi _{0}}{m_{\xi }^{2}}\,(\frac{T}{T_{0}})^{3/2}
\equiv \theta _{0}\,(\frac{T}{T_{0}})^{3/2} \,, \hspace{0.5cm}
h = (\frac{2\omega }{m_{\xi }})^{2} \equiv  n^{2} + \epsilon \,.
\end{eqnarray}
The initial $\theta $ value,
\( \:
\theta _{0} = \frac{2\mu \xi _{0}}{m_{\xi }^{2}} \,,
\: \)
is the most crucial quantity for the cosmological fate of $\xi $ field.

Effect of the cosmological redshift is very important in the case of
small amplitude perturbation of $2|\theta | \leq h$, because it may readily
push away resonance frequencies beyond the narrow band.
If a frequency redshifts outside the band, the instability of this
particular mode ceases to operate. Under the small amplitude perturbation
the time $\Delta t$ during which a particular frequency
stays within the $n-$th band is determined by
\begin{eqnarray}
1 - \frac{a(t)}{a(t+\Delta t)}
< \Delta \epsilon_{n} \,.
\end{eqnarray}
Assuming a small $\Delta \epsilon_{n} $, one derives
\begin{eqnarray}
\Delta t\,H < \Delta \epsilon_{n} \,.
\end{eqnarray}
With
\( \:
\Delta \epsilon _{n} \ll 1 \,,
\: \)
the walk-out time is much smaller than the Hubble time
$H^{-\,1}$.

At the walk-out time
\( \:
\Delta t_{R} = \frac{\Delta \epsilon_{n} }{H} \,,
\: \)
the exponent factor in the produced number density is
\begin{eqnarray}
\gamma _{n}\,\Delta t_{R} = \frac{(\Delta \epsilon _{n})^{2}}{4n}\,
\frac{m_{\xi }}{H} =
\frac{\theta _{0}^{2n}}{2^{4n-4}\,n\,[(n-1)!]^{4}}\,(\frac{T}{T_{0}})^{3n-2}
\,,
\end{eqnarray}
assuming the temperature dependence of
the Hubble parameter in the radiation dominated epoch:
\( \:
H \:\propto  \: T^{2} \,.
\: \)
Growth rate $R_{n}$ while staying within the band is then given by
dividing the number density of produced particles by the number density
of parent $\xi $ particles, $\frac{1}{2}\, m_{\xi }\,\xi ^{2}$.
By taking one species of a massless particle for simplicity,
the  production rate reads as
\begin{eqnarray}
R_{n} = \frac{n^{3/2}}{2^{5}\,\sqrt{2}\,\pi ^{3/2}}
\,(\frac{m_{\xi }}{\xi _{0}})^{2}\,(\frac{T}{T_{0}})^{-\,2}\,
\exp [\,C_{n}\,(\frac{T}{T_{0}})^{3n-2}\,] \,, \hspace{0.5cm}
C_{n} = \frac{16\,(\mu \xi _{0}/m_{\xi }^{2})^{2n}}{2^{2n}\,n\,[(n-1)!]^{4}}
\,.
\end{eqnarray}
The ratio of produced energy density to the parent's is
\begin{eqnarray}
\frac{\rho }{\rho _{\xi }} = \frac{n}{2}\,R_{n} \,.
\end{eqnarray}
In particular, the dominant fundamental mode of $n=1$ contributes with
the rate,
\begin{eqnarray}
R_{1} \simeq  4.0 \times 10^{-3}
\,(\frac{m_{\xi }}{\xi _{0}})^{2}\,(\frac{T}{T_{0}})^{-\,2}\,
\exp [\,4\,(\frac{\mu \xi _{0}}{m_{\xi }^{2}})^{2}\,\frac{T}{T_{0}}\,] \,.
\label{small rate}
\end{eqnarray}
This formula is valid only in the radiation dominated epoch. In the
$\xi -$matter dominated epoch the temperature dependence in the exponent
is modified to $(T/T_{0})^{3/2}$.
The prefactor
\( \:
(\frac{T}{T_{0}})^{-\,2}
\: \)
increases as time proceeds, hence as $T$ decreases. This cannot be true
for an indefinitely long time.
Indeed, our asymptotic formula of produced number
of particles is valid only for
\( \:
\frac{1}{2}\, \lambda _{n} m_{\xi }\,t \geq 1 \,,
\: \)
that is, when the exponent above is large, hence when
\( \:
\frac{T}{T_{0}}
\: \) is not too small.
Here the strong coupling or the large amplitude problem is involved.

The production rate under the large amplitude perturbation
depends on various factors in different ways from the small amplitude case,
in particular on the oscillation amplitude $\xi $, which may drastically
change the fate of WISF.
The frequency range that contributes to the growing particle production
depends both on the initial $\theta _{0}$ value and on the temperature
scale in question $T$, relative to $T_{0}$.
For simplicity we shall only consider a moderate range of the initial
parameter
\( \:
\theta _{0} = \frac{2\mu \xi _{0}}{m_{\xi }^{2}} \,,
\: \)
for instance, $O[20]$ or even less,
which turns out to be sufficient for later application.
The large amplitude perturbation region available
is then restricted, initially to a wide $h<2\theta $ region and at later
stages to a few adiabatic regions, Region \lromn 1,
Region \lromn 2, Region \lromn 3, etc., as explained in
Section \lromn 4.C.
One must check in each case of application whether the necessary
condition of applicability is fulfilled or not.
Before going into any details of how this is met,
we shall give an order of magnitude estimate
of production rate in the adiabatic region.
The number of produced particles of one light species of particles with
\( \:
m \ll \sqrt{\mu \xi _{0}}\,(\frac{T}{T_{0}})^{3/4}
\: \)
up to one Hubble time is
\begin{eqnarray}
&& \hspace*{-1cm}
R = \frac{\langle \,N\,\rangle /V}{m_{\xi }\xi ^{2}/2} \simeq
2.0 \times 10^{-2}\,
\frac{\mu ^{3/2}}{m_{\xi }\,\xi _{0}^{1/2}}\,(\frac{T}{T_{0}})^{-\,3/4}\,
\exp [\,(\,2.4 - 0.65\,)\,\sqrt{\frac{\mu \xi_{0} }{m_{\xi }^{2}}}
\,(\frac{T_{0}}{T})^{5/4}\,] \,, \\
&& \hspace*{-1cm}
\frac{\rho }{\rho _{\xi }} = 1.5 \times 10^{-2}\,(\frac{\mu }{m_{\xi }})^{2}\,
\exp [\,(\,2.4 - 0.65\,)\,\sqrt{\frac{\mu \xi_{0} }{m_{\xi }^{2}}}
\,(\frac{T_{0}}{T})^{5/4}\,] \,,
\end{eqnarray}
extending the phase space region to the full
\( \:
0 < \omega < \sqrt{\mu \xi } \,.
\: \)
Two numbers written in the exponent correspond to the adiabatic
approximation with and without the $\omega $ dependent reduction factor
of the rate coefficient.
The available phase space is actually reduced from this one, by a reduction
factor of the form,
\( \:
f[\,\frac{2\mu \xi _{0}}{m_{\xi }^{2}}\,(\frac{T}{T_{0}})^{3/2}\,] \,.
\: \)
A precise form of the function $f(x)$ will be given below when
a specific application is made.
The crucial exponent factor increases like
\( \:
(T_{0}/T)^{5/4}
\: \)
as time proceeds, in contrast to the
 $T/T_{0}$ factor in the small amplitude case
Eq.\ref{small rate}.
The precise temperature scale dependence
\( \:
(T_{0}/T)^{5/4}
\: \)
here is trusted only when the coupling
factor $\mu $ is independent of the particle energy $\omega $, which does not
hold in the axion case having a different $(T_{0}/T)^{1/2}$ behavior.

The production rate given here is a yield produced during
one Hubble time after the epoch of temperature $T$.
There always exists accumulation effect of produced particles, but
with a rapidly changing production rate the main contribution comes from
one Hubble time at the maximal production. We therefore ignore a minor
accumulation effect.

With these in mind we proceed to discuss important cases of application.
First, the invisible axion. The axion field which we denote by $A$
has an effective two photon coupling in the form,
\begin{eqnarray}
{\cal L}_{{\rm eff}} =
\frac{2\alpha }{\pi }\,\frac{1}{f_{A}}\,A\,\vec{E}\cdot \vec{B} \,.
\end{eqnarray}
The fine structure constant $\alpha $ appears, because this term is
induced by a triangular loop of fermions. $f_{A}$ is the scale at which
the Peccei-Quinn symmetry is broken and
we ignored $O[1]$ factors in the coupling.
The other coupling to matter fields such as quarks and leptons
are irrelevant to our problem.
The effective Lagrangian yields a modified Maxwell equation,
\begin{eqnarray}
\dot{\vec{E}} - \vec{\nabla} \times \vec{B} =
\frac{\alpha }{\pi }\,\frac{1}{f_{A}}\,\dot{A}\,\vec{B}
\,.
\end{eqnarray}
Relevant parameters for the resonant production is
\begin{eqnarray}
\mu  \simeq  \frac{\alpha }{\pi }\,\frac{m_{A}}{f_{A}}\,\omega
\,, \hspace{0.5cm}
\xi _{0} = f_{A} \,, \hspace{0.5cm}
m_{A} \sim \frac{\Lambda ^{2}}{f_{A}} \,,
\end{eqnarray}
with $\Lambda $ the QCD scale.
The quantity $\theta _{0}$ is dependent on the photon energy $\omega $ like
\begin{eqnarray}
\theta _{0} = \frac{2\mu \xi _{0}}{m_{A}^{2}} \sim
\frac{2\alpha }{\pi }\,\frac{\omega }{m_{A}} \,.
\end{eqnarray}
This factor is too small for
\( \:
\frac{\omega }{m_{A}} \ll \pi /(2\alpha)
\: \),
and no appreciable
resonant production occurs for the axion as far as the small amplitude
formula Eq.\ref{small rate} can be applied,
because frequencies immediately
redshift away outside the band  by the time
$\Delta \epsilon \,H^{-\,1} \ll H^{-\,1}$ \cite{parametric redshift}.

Now the important question is whether there exists a region of
parameters in which the large amplitude formula can be applied.
This cannot occur in the axion case, because
the stability gap constraint applied to the axion leads to
incompatible conditions like
\begin{eqnarray}
\omega < \frac{\alpha }{\pi }\,m_{A}\,(\frac{T}{T_{0}})^{3/2}
\,, \hspace{0.5cm}
\omega > 0.16 \,m_{A}\,
[\,\frac{\alpha }{\pi }\,(\frac{T}{T_{0}})^{3/2}\,]^{-1} \,. \nonumber
\end{eqnarray}
Basically the presence of the stability triangle restricts
\( \:
\theta \geq 0.33 \,,
\: \)
which has no overlap with the adiabatic $h<2\theta $ condition due to
the small $\frac{\alpha }{\pi }$ factor.
Thus there is no way an appreciable resonant production takes place for
the cosmological axion.

The situation is different for fields of the Polonyi type.
Here a generic dimensionless coupling may be of order unity or larger,
but the effective
coupling may be suppressed by a small ratio of an intermediate to a large
mass scale.
We express relevant parameters by
\begin{eqnarray}
m_{\xi } = \frac{F^{2}}{M} \,, \hspace{0.5cm}
\xi _{0} = M \,,
\hspace{0.5cm}
\mu  = g_{0}\,\frac{F^{4}}{M^{3}} \,,
\end{eqnarray}
leading to the initial amplitude,
\begin{eqnarray}
\theta _{0} = 2g_{0} \,.
\end{eqnarray}
The large mass scale $M$ is the Planck scale and the intermediate
mass scale $F$ is of order $10^{10}$ GeV or more
in the case of the Polonyi field.
We only consider the Yukawa coupling to spinless bosons, but there may
exist coupling to the gauge field similar to the two photon decay of
the axion, in which case
the following estimate is modified in substantial ways.
Assumption of the small amplitude oscillation leads to
the growth rate while staying within the resonance band,
\begin{eqnarray}
\frac{\rho _{1}}{\rho _{\xi }} = 0.99 \times 10^{-3}\,(\frac{F}{M})^{4}\,
(\frac{T}{T_{0}})^{-\,2}\,
\exp [\,4g_{0}^{2}\,\frac{T}{T_{0}}\,] \,.
\end{eqnarray}
Unless $g_{0}^{2}$ is very large, for instance $250$ or more, it is again
unlikely that this type of WISF decays in the small amplitude regime.

One must however use formulas relevant to the large amplitude perturbation
if the exponent factor is large.
The large amplitude formula yields production rate
of particles in the mass range of
\( \:
m \ll  \sqrt{g_{0}}\,m_{\xi }\,(\frac{T}{T_{0}})^{3/4} \,,
\: \)
\begin{eqnarray}
\frac{\rho }{\rho _{\xi }} \simeq
1.5 \times 10^{-2}\,g_{0}^{2}\,(\frac{F}{M})^{4}\,
f[\,2g_{0}(\frac{T}{T_{0}})^{3/2}\,]\,
\,\exp [\,(\,2.4 - 0.65\,)\,\sqrt{g_{0}}\,(\frac{T_{0}}{T})^{5/4}\,] \,,
\end{eqnarray}
in the radiation dominated epoch.
The two numbers in the exponent, $2.4$ and $0.65$, correspond to the
adiabatic approximation with and without the $\omega -$dependent reduction
factor of the rate coefficient.
A more refined estimate would lead to a modified exponent
replacing the coefficient $(\,2.4 - 0.65\,)$,
which may weakly depend on
\( \:
g_{0}\,(T/T_{0})^{3/2} \,.
\: \)
$f(x)$ is the reduction factor of the phase space due to the restricted
$\omega -$range in the adiabatic region, which is given in
regions of immediate interest,
\begin{eqnarray}
f(x) &\sim & 1 \,, \hspace{0.5cm} (x \geq O[10]) \,,
\nonumber \\
&=& 1 - \frac{1}{4}\,(\frac{5}{x} + 1 - \frac{5}{24}\,x)\,
(\frac{3}{x} - 1 + \frac{x}{24})\,,  \hspace{0.5cm}
(1.9 < x < 3) \,, \nonumber \\
&=& (\frac{1}{2x} + \frac{1}{2} - \frac{x}{16})^{2} \,,  \hspace{0.5cm}
(0.90 < x < 1.9) \,, \nonumber \\
&=& (\frac{1}{2} + \frac{1}{2x} - \frac{x}{16})\,
(\frac{3}{2} - \frac{1}{2x} + \frac{x}{16}) \,, \hspace{0.5cm}
(0.33 < x < 0.90)
\,.
\end{eqnarray}
Using these functional forms, one can readily find the temperature scale
that the produced particle energy becomes comparable to the parent's.
We did some sample computations. For instance, for
\( \:
g_{0} = 15
\: \)
we find that the decay occurs at
\begin{eqnarray}
\frac{T}{T_{0}} \sim 0.2 \,,
\label{polonyi decay time}
\end{eqnarray}
taking
\( \:
F/M = 10^{-9} - 10^{-6}
\: \)
with the larger exponent $2.4$, and
\begin{eqnarray}
\frac{T}{T_{0}} \sim 0.06 - 0.09 \,,
\end{eqnarray}
with
the smaller exponent $0.65$.

For a smaller $g_{0}$ value one must take into account of
the adiabatic constraint of $\theta > 0.33$, since no sizable particle
production is expected in the small amplitude regime.
Ignoring the phase space suppression factor of order unity,
the produced energy density roughly obeys
\begin{eqnarray}
\frac{\rho }{\rho _{\xi }} \approx  1.5 \times 10^{-2}\,g_{0}^{2}\,
(\frac{F}{M})^{4}\,\exp \,[\,(\,11 - 3.0\,)\,g_{0}^{4/3}\,] \,,
\end{eqnarray}
right at $\theta = 0.33$ in the radiation dominated epoch.
For this quantity  to exceed $1$ so that the decay is completed in the
large amplitude regime, the coupling $g_{0}$ must satisfy
\( \:
g_{0} > 4 - 3
\: \)
with the larger exponent $11$, and
\( \:
g_{0} > 12 - 9
\: \)
with
the smaller exponent $3.0$, taking for the intermediate mass scale
\( \:
F/M = 10^{-9} - 10^{-6} \,.
\: \)
Thus there seems no fundamental difficulty in getting a substantial decay
of the Polonyi type field, provided that $g_{0} \geq  O[10]$.
The precise number of this bound needs to be calculated by
more sophisticated ways, but it should be clear that there is a definite
possibility of resolving the Polonyi problem.

One possible problem concerns effect of the Polonyi mass dominance on
the particle production rate.
Modified time variation of the decayed energy density during the $\xi -$matter
dominated epoch is given by
\begin{eqnarray}
\frac{\rho }{\rho _{\xi }} =
1.5 \times 10^{-2}\,g_{0}^{2}\,(\frac{F}{M})^{4}\,
f[\,2g_{0}(\frac{T}{T_{0}})^{3/2}\,]\,
\,\exp [\,(\,2.4 - 0.65\,)\,\sqrt{g_{0}}\,(\frac{T_{0}}{T_{m}})^{5/4}\,
(\frac{T_{m}}{T})^{3/4}\,] \,,
\end{eqnarray}
with $T_{m}$ the temperature at the onset of $\xi -$matter dominance,
\begin{eqnarray}
\frac{T_{m}}{T_{0}} = \frac{4\pi }{3}\,(\frac{M}{m_{\rm pl}})^{2} \,.
\end{eqnarray}
The rate in the matter dominated epoch increases more slowly than in
the radiation dominated epoch, but still rapidly enough
as the temperature $T$ decreases.
Right at the onset of matter dominance of $T = T_{m}$
the rate is given by
\begin{eqnarray}
&&
\left( \frac{\rho }{\rho _{\xi }}\right)_{{\rm m}} =
1.5 \times 10^{-2}\,g_{0}^{2}\,(\frac{F}{M})^{4}\,
f[\,2g_{0}\,(\frac{4\pi }{3})^{3/2}\,(\frac{M}{m_{\rm pl}})^{3}\,]
\nonumber \\
&& \hspace*{2cm}
\cdot \exp [\,(\,2.4 - 0.65\,)\,\sqrt{g_{0}}\,(\frac{3}{4\pi })^{5/4}\,
(\frac{m_{{\rm pl}}}{M})^{5/2}\,] \,.
\end{eqnarray}
Taking
\( \:
M = m_{{\rm pl}}/\sqrt{8\pi }
\: \)
as an illustration
gives the exponent factor as large as
\( \:
= 10^{\,(\,23 - 6.2\,)\,\sqrt{g_{0}}} \,.
\: \)

In order to complete the Polonyi decay at a temperature scale
prior to the matter dominated epoch, there is a condition on $\xi $
parameters.
These come from $\theta > 0.33$ and
\( \:
\left( \frac{\rho }{\rho _{\xi }}\right)_{{\rm m}} > 1 \,,
\: \)
leading to
\begin{eqnarray}
&&
g_{0}\,(\frac{M}{m_{{\rm pl}}})^{3} > 0.019 \,, \\
&& \hspace*{-1cm}
1.5 \times 10^{-2}\,g_{0}^{2}\,(\frac{F}{M})^{4}\,
f[\,17\,g_{0}\,(\frac{M}{m_{\rm pl}})^{3}\,]\,
\,\exp [\,(\,0.40 - 0.11\,)\,\sqrt{g_{0}}\,
(\frac{m_{{\rm pl}}}{M})^{5/2}\,] > 1 \,.
\end{eqnarray}
Again taking
\( \:
M = m_{{\rm pl}}/\sqrt{8\pi }
\: \)
as a guide and ignoring the phase space reduction factor of order unity,
we find a constraint,
\begin{eqnarray}
1.5 \times 10^{-2}\,g_{0}^{2}\,(\frac{F}{M})^{4}\,
\exp [\,(\,23 - 6.2\,)\,\sqrt{g_{0}}\,] > 1 \,,
\end{eqnarray}
with $g_{0} > 2.4$.
In the range of the intermediate mass scale of
\( \:
F/M = 10^{-9} - 10^{-6} \,,
\: \)
this constraint gives
\( \:
g_{0} > 13 - 6.1
\: \)
for the larger exponent factor of $23$ and
\( \:
g_{0} > 160 - 69
\: \)
for the smaller exponent factor of $6.2$.

Although these numbers for the coupling seem a bit large,
we note that the Polonyi field does not have to decay completely
prior to the epoch of its dominance. Even in the $\xi -$matter dominated
epoch the particle production rate increases, albeit with a slower rate
than in the radiation dominated epoch.
Right at the adiabatic limit of $\theta = 0.33$ the decayed energy
density in the $\xi -$dominated epoch becomes
\begin{eqnarray}
\frac{\rho }{\rho _{\xi }} \approx 1.5 \times 10^{-2}\,g_{0}^{2}\,
(\frac{F}{M})^{4}\,\exp \,[\,(\,2.9 - 0.79\,)
g_{0}\,\frac{m_{{\rm pl}}}{M}\,]\,,
\end{eqnarray}
again ignoring the phase space reduction factor of order unity.
In order to get the Polonyi decay within the large amplitude regime,
the right hand side quantity must exceed unity. With
$M = m_{{\rm pl}}/\sqrt{8\pi }$ as a guide, this happens for the larger
exponent of $2.9$ at $g_{0} = 5.5 - 3.6$, and for the smaller exponent
of $0.79$ at $g_{0} = 19 - 13$, taking the mass range of
\( \:
F/M = 10^{-9} - 10^{-6} \,.
\: \)

A possible problem that might arise when the Polonyi field decays in
the epoch of its dominance is a large entropy creation after
the Polonyi energy is completely converted to radiation.
This might dilute the baryon asymmetry created at the GUT epoch.
We may estimate the entropy dilution factor $S$ by assuming instantaneous
conversion of the Polonyi energy into thermal radiation.
The reheated thermal temperature $T_{R}$ is determined by
\begin{eqnarray}
d_{R}\,T_{R}^{4} = d_{m}\,T_{m}^{4}\,(\frac{T}{T_{m}})^{3} \,,
\end{eqnarray}
where the temperature $T$ is obtainted by solving
\( \:
\frac{\rho }{\rho _{\xi }} = 1 \,.
\: \)
Ignoring a difference in the massless degrees of freedom $d_{i}$ in
different epoch, one finds for the entropy dilution factor,
\begin{eqnarray}
S \sim  (\frac{T_{R}}{T})^{3} \sim  (\frac{T_{m}}{T})^{3/4} \,.
\end{eqnarray}
The further T deviates from $T_{m}$, the more entropy is created,
but with the dilution factor governed
by a not too large exponent factor of $3/4$.
With
\( \:
M = m_{{\rm pl}}/\sqrt{8\pi }
\: \)
the ratio of produced energy density to the parent's becomes
\begin{eqnarray}
\frac{\rho }{\rho _{\xi }} = 1.5 \times 10^{-2}\,g_{0}^{2}\,(\frac{F}{M})^{4}
\,\exp \,[\,(\,23 - 6.2\,)\,\sqrt{g_{0}}\,(\frac{T_{m}}{T})^{3/4}\,] \,,
\end{eqnarray}
under the large amplitude constraint
\( \:
g_{0}\,(\frac{T}{T_{m}})^{3/2} > 2.4 \,.
\: \)
One may thus compute the dilution factor $S(g_{0})$ for $g_{0} > 2.4$.
For instance, with $g_{0} = 10 - 50$ the entropy dilution factor is computed
as
\( \:
S \sim 4 - 2
\: \)
at most, which may be regarded as a minor effect.
Given many uncertainties on the Polonyi field, especially on values of
\( \:
g_{0} \; {\rm and} \; M \,,
\: \)
we believe the situation promissing for the resolution of the Polonyi
problem.

One great uncertainty not considered here is that
the Polonyi mass may not be as simple as a constant $m_{\xi }$.
In general the Polonyi mass
may depend on the field amplitude $\xi $.
Furthermore, the Polonyi field may have decay channels other than
spinless bosons, in particular massless gauge bosons like the photon
and the gluon. In this case our estimate will be modified.
For instance, if the Polonyi field decays dominantly into massless gauge
boson pairs, then the exponential growth terminates at the onset of the
Polonyi mass dominance in the case of $g_{0}$ not large enough,
and the ordinary decay at a much later epoch
might create too much entropy.
A more detailed analysis based on particular models
is called for, but the catastrophic decay discussed here may
potentially give a great
impact on phenomenological model building of supergravity models.

\vspace{2cm}
{\bf Acknowledgement}

\vspace{0.5cm}
The author should like to thank
M.Yamaguchi for a few critical comments on cosmological application
and for a discussion on the Polonyi field, and thank
T.Goto
for a discussion on the Polonyi field.

%%%%%%%%%%%%%%%%%%%%%%%%%%%%%%%%%%%%%%%%%%%%%%%

\newpage
{\bf Appendix
\hspace{0.3cm}
Parametric resonance in the weak coupling}

\vspace{0.5cm}

\vspace{0.5cm}
We follow in spirit the physicist approach with guesswork and consistency
check of results, by generalizing analysis of Ref.\cite{landau-lifschitz m}.
A Floquet type of solution is thus expanded as
\begin{eqnarray}
&& \hspace*{-2.5cm}
e^{\lambda  z}\,P(z)  =
\sum_{k = -\,\infty }^{\infty }\,c_{k}(z)\,e^{i(n+2k)z} \,,
\nonumber
\end{eqnarray}
with the periodicity understood:
\( \:
P(z+\pi ) = P(z) \,.
\: \)
Here we focus on the weak coupling or the small amplitude
case, thus the instability region is expected to be in a narrow band.

The basic idea of our approach is that within the narrow band near
\( \:
h = n^{2}
\: \)
one may take $e^{\pm inz}$ as the zero-th order
approximate solution and assume
small variation of the amplitude $c_{k}(z)$. It is thus legitimate to
keep time derivative terms only in the two modes,
\( \:
k = 0 \; {\rm and} -n \,,
\: \)
in the form of
\( \:
\dot{c}_{0 \,,-\,n} \sim \lambda  c_{0 \,,-\,n} \,,
\hspace{0.3cm} {\rm since} \;
|\lambda  | \ll 1
\,.
\: \)
We anticipate a systematic hierarchy of coefficients in the small
parameter $\sqrt{\theta _{+}\theta _{-}}$:
\begin{eqnarray*}
c_{k} = O[\,(\theta _{+}\theta _{-})^{{\rm Min.}\;(\,|k|/2 \,, |k+n|/2\,)}\,]
\,.
\end{eqnarray*}
With
\( \:
h = n^{2} + \epsilon \,, \; |\epsilon | \ll 1 \,,
\: \)
the original equation leads to an infinite set of relations among coefficients,
\begin{eqnarray}
&&
\gamma _{k}\,c_{k} - \theta _{+}\,c_{k+1} - \theta _{-}\,c_{k-1} = 0 \,,
\\
&& \hspace*{-1cm}
\gamma _{k} =  -\,4k(n+k) \hspace{0.3cm}
(k \neq 0 \,, -\,n) \,, \hspace{0.3cm}
\gamma _{0} = \epsilon + 2in\lambda  \,, \;
\gamma _{-\,n} = \epsilon - 2in\lambda
\,.
\end{eqnarray}
In matrix notation
\begin{eqnarray}
\left(
\begin{array}{ccccccccc}
0 & \cdots & \cdots & \cdots & \cdots & \cdots & \cdots & \cdots & 0
\\
0 & \cdots  & -\,\theta _{+} & \gamma _{k} & - \,\theta _{-} & &
& \cdots &  0 \\
0 & \cdots & \cdots & \cdots & \cdots & \cdots & \cdots & \cdots & 0
\\
0 & \cdots &\cdots  & \cdots & -\,\theta _{+} & \gamma _{-\,k}
& - \,\gamma _{-}
& \cdots & 0 \\
0 & \cdots & \cdots & \cdots & \cdots & \cdots & \cdots & \cdots & 0
\end{array}
\right)
\,
\left(
\begin{array}{c}
\cdot   \\
c_{k} \\
\cdot \\
c_{-\,k} \\
\cdot
\end{array}
\right)
= 0 \,.
\end{eqnarray}

To proceed further, it is useful to separately treat
the three blocks defined by
\( \:
k \geq 1 \,, \; k \leq -\,n - 1 \,, \; -\,n + 1 \leq k \leq  -\,1 \,,
\: \)
and leave to the last
the main
\( \:
c_{0} \;(k=0)\,, \;c_{-\,n}\;(k= -\,n)
\: \)
equation.
Noting that
\( \:
\gamma _{k} = \gamma _{-\,n-k} \,,
\: \)
one finds that the blocks with
\( \:
k \geq 1 \; {\rm and} \; k \leq -\,n - 1
\: \)
yield identical results and
\begin{eqnarray}
&&
c_{1} = \theta _{-}\,c_{0}\,\lim_{N \rightarrow \infty }\,(D_{N}^{-\,1})_{11}
\,, \hspace{0.5cm}
c_{-\,n-1} =
\theta _{+}\,c_{-\,n}\,\lim_{N \rightarrow \infty }\,(D_{N}^{-\,1})_{11}
\,, \\
&& \hspace*{-1cm}
D_{N} =
\left(
\begin{array}{cccccc}
\gamma _{N} & -\,\theta _{-} & 0 & \cdots & \cdots & 0  \\
-\,\theta _{+} & \gamma _{N-1} & -\,\theta _{-} & 0 & \cdots & 0 \\
0 & \cdots & \cdots & \cdots & \cdots & 0 \\
0 & \cdots & \cdots & \cdots & \cdots & 0 \\
0 & \cdots  & 0 & -\,\theta _{+} & \gamma _{2} & -\,\theta _{-} \\
0 & \cdots & \cdots  & 0 & -\,\theta _{+} & \gamma _{1}
\end{array}
\right) \,.
\end{eqnarray}
The central block of
\( \:
-\,n + 1 \leq k \leq  -\,1
\: \)
is then solved for
\( \:
c_{-\,1} \; {\rm and} \; c_{-\,n+1}
\: \)
in terms of
\( \:
c_{0} \; {\rm and} \; c_{-\,n} \,,
\: \)
yielding
\begin{eqnarray}
&&
c_{-\,1} =
[\:(E^{-\,1})_{11}\,\theta _{+}\,
c_{0} + (E^{-\,1})_{1\,, n-1}\,\theta _{-}\,c_{-\,n} \:]
\,,  \\
&&
c_{-\,n + 1} = [\:(E^{-\,1})_{n-1\,, 1}\,
\theta _{+}\,c_{0} + (E^{-\,1})_{n-1\,, n-1}\,\theta _{-}\,c_{-\,n} \:]
\,,
\\
&& \hspace*{-1cm}
{\rm with} \hspace{0.5cm}
E =
\left(
\begin{array}{cccccc}
\gamma _{-\,1} & -\,\theta _{-} & 0 & \cdots & \cdots & 0  \\
-\,\theta _{+} & \gamma _{-\,2} & -\,\theta _{-} & 0 & \cdots & 0 \\
0 & \cdots & \cdots & \cdots & \cdots & 0 \\
0 & \cdots & \cdots & \cdots & \cdots & 0 \\
0 & \cdots  & 0 & -\,\theta _{+} & \gamma _{-\,n+2} & -\,\theta _{-} \\
0 & \cdots & \cdots  & 0 & -\,\theta _{+} & \gamma _{-\,n+1}
\end{array}
\right) \,.
\end{eqnarray}

Finally, we insert back $c_{\pm 1}$ and $c_{-\,n \pm 1}$ into the last
\( \:
c_{0} \;(k=0)\,, \;c_{-\,n}\;(k= -\,n)
\: \)
equation.
Noting that
\( \:
E^{-\,1}_{11} = E^{-\,1}_{n-1\,, n-1} \,,
\: \)
one finds that
\begin{eqnarray}
&&
\hspace*{-2.5cm}
\left(
\begin{array}{cc}
\epsilon + 2in\lambda  -
\theta _{+}\theta _{-}\,(D^{-\,1}_{11} + E^{-\,1}_{11})
&  -\,\theta _{-}^{2}\,
E^{-\,1}_{1\,, n-1}
\\
-\,\theta _{+}^{2}\,E^{-\,1}_{n-1\,, 1} &
\epsilon - 2in\lambda  - \theta _{+}\theta _{-}\,
(D^{-\,1}_{11} + E^{-\,1}_{11})
\end{array}
\right)
\,
\left(
\begin{array}{c}
c_{0}  \\
c_{-\,n}
\end{array}
\right) = 0  \,.
\end{eqnarray}
Nontrivial solution exists only for
\begin{eqnarray}
&& \hspace*{-1cm}
(2n\lambda  )^{2} = (\theta _{+}\theta _{-})^{2}\,
E^{-\,1}_{n-1\,, 1}\,E^{-\,1}_{1\,, n-1} -
[\: \epsilon - \theta _{+}\theta _{-}\,(D^{-\,1}_{11} + E^{-\,1}_{11} )\:]^{2}
\equiv (\epsilon _{+} - \epsilon )\,(\epsilon  - \epsilon _{-})\,,
 \nonumber \\
&&
\\
&&
\epsilon _{\pm } = \theta _{+}\theta _{-}\,
(D^{-\,1}_{11} + E^{-\,1}_{11} )
\pm \theta _{+}\theta _{-}\,
\sqrt{\: E^{-\,1}_{n-1\,, 1}\,E^{-\,1}_{1\,, n-1}\:} \,.
\end{eqnarray}

We then compute the inverse matrix elements to lowest non-trivial orders.
We thus find that
\begin{eqnarray}
&& \hspace*{0.5cm}
\epsilon _{\pm }
\simeq
\frac{1}{2}\, \frac{1}{n^{2}-1} \,\theta _{+}\theta _{-} \pm
\frac{(\theta _{+}\theta _{-})^{n/2}}
{(\:2^{n-1}\,(n-1)! \:)^{2}}
\,,\hspace{0.3cm}
(n\neq 1) \,.
\end{eqnarray}
Thus the band width of parametric resonance is given by
\begin{eqnarray}
\Delta \epsilon =
\frac{(\theta _{+}\theta _{-})^{n/2}}
{2^{2n-3}\,(\,(n-1)! \,)^{2}} \,.
\end{eqnarray}
The growth rate $\lambda  $ takes a simple form,
\begin{eqnarray}
\lambda  (\epsilon ) = \frac{1}{2n}\,\sqrt{\: (\frac{\Delta \epsilon }{2})^{2}
- (\,\epsilon - \frac{\epsilon _{+} + \epsilon _{-}}{2} \,)^{2} \:}
\,.
\end{eqnarray}

The two special cases of $n = 0$ and $n=1$
must be dealt with separately. First, $n= 1$ case.
By using
\begin{eqnarray}
\hspace*{-1.5cm}
\left(
\begin{array}{cc}
\epsilon + 2in\lambda  -  \theta _{+}\theta _{-}\,(D^{-\,1})_{11} &
-\,\theta _{-}
 \\
-\,\theta _{+} &
\epsilon - 2in\lambda  -  \theta _{+}\theta _{-}\,(D^{-\,1})_{11}
\end{array}
\right) \,
\left(
\begin{array}{c}
c_{0}  \\
c_{-\,1}
\end{array}
\right) = 0 \,,
\end{eqnarray}
we obtain
\begin{eqnarray}
\epsilon _{\pm } = \theta _{+}\theta _{-}\,(D^{-\,1})_{11} \pm
\sqrt{\: \theta _{+}\theta _{-} \:} \simeq
-\,\frac{1}{8}\,\theta _{+}\theta _{-} \pm \sqrt{\: \theta _{+}\theta _{-} \:}
\,,
\end{eqnarray}
with
\( \:
D \equiv \lim_{N \rightarrow \infty }\,D_{N} \,.
\: \)

Next, we deal with $n=0$ case. In this case there is no linear term of
$\dot{c}_{0}$, and one must retain the higher order term of
$\stackrel{..}{c}_{0}$.
Moreover, there is no intermediate block,
and the final coeffficient equation reads as
\begin{eqnarray}
&&
-\,\theta _{+}\,c_{1} + \gamma _{0}\,c_{0} - \theta _{-}\,c_{-1} = 0
\,, \hspace{0.5cm}
\gamma _{0} = \lambda ^{2} + \epsilon \,, \\
&& \hspace*{-2cm}
{\rm leading \; to} \hspace{0.2cm}
\lambda = \sqrt{\,-\,\frac{1}{4}\,\theta _{+}\theta _{-} - \epsilon \,}
\,, \hspace{0.5cm}
\epsilon < -\,\frac{1}{4}\,\theta _{+}\theta _{-} \,.
\end{eqnarray}
Thus, there is no band of $n=0$, and a gap extending to $n=1$ band exists in
the region,
\begin{eqnarray}
-\,\frac{1}{4}\,\theta _{+}\theta _{-} < h < 1
- \sqrt{\,\theta _{+}\theta _{-}\,} - \frac{1}{8}\,\theta _{+}\theta _{-}
\,.
\end{eqnarray}

It is also possible to compute coefficients in terms of an overall
factor $c_{0}$.
We quote the result,
\begin{eqnarray}
&&
\frac{c_{0}}{c_{-\,n}} = (\frac{\theta _{-}}{\theta _{+}})^{n/2}\,
e^{-\,2i\eta } \,, \hspace{0.5cm}
\tan \eta  =
\sqrt{\:\frac{\epsilon _{+} - \epsilon }{\epsilon - \epsilon _{-}} \:}
\,, \\
&&
c_{-\,k} =
(\:E^{-\,1}_{k1}\,\theta _{+}\,c_{0} + E^{-\,1}_{k\,,n-1}
\,\theta _{-}\,c_{-\,n}\:)
\nonumber \\
&& \hspace*{-2cm}
= \frac{\theta _{+}^{k/2}\,\theta _{-}^{(n-k)/2}}
{\sqrt{\:\theta _{+}^{n
} + \theta _{-}^{n}\:}}\,
[\:
\frac{(\theta _{+}\theta _{-})^{k/2}\,e^{-\,i\eta }}
{2^{2(k-1)}\,[(k-1)!]^{2}} +
\frac{(k!)^{2}\,(\theta _{+}\theta _{-})^{(n-k)/2}\,e^{i\eta }}
{2^{2(n-k-1)}\,[(n-1)!]^{2}}
\:]\,,  \hspace{0.3cm} (1 \leq k \leq n-1) \,, \\
&&
c_{k}
\simeq
(-\,1)^{k}\,\frac{n!}{2^{2k}\,k!\,(n+k)!}\,\theta _{-}^{k}\,c_{0}
\,,
\hspace{0.3cm} (k\geq 1) \,,
\nonumber \\
&&
c_{-\,n -k} \simeq
(-\,1)^{k}\,\frac{n!}{2^{2k}\,k!\,(n+k)!}\,\theta _{+}^{k}\,c_{-\,n} \,,
\hspace{0.3cm} (k\geq 1) \,.
\end{eqnarray}
As promised, the expansion coefficients of terms
\( \:
e^{\pm inz + ikz}
\: \)
systematically contain suppression factors of
\( \:
\theta ^{|k|} \,, \; \theta \equiv \sqrt{\,|\theta _{+}\,\theta _{+}\,|} \,.
\: \)
To leading order the functional form in the $n-$th band is then
\begin{eqnarray}
&& \hspace*{-2cm}
u
\sim
c_{0}\,e^{\lambda z + inz} + c_{-\,n}\,e^{\lambda z - inz}
=
\frac{\theta _{-}^{n/2}}{\sqrt{\:\theta _{+}^{n} + \theta _{-}^{n}\:}}\,
e^{\lambda  z + i(nz - \eta )} +
\frac{\theta _{+}^{n/2}}{\sqrt{\:\theta _{+}^{n} + \theta _{-}^{n}\:}}\,
e^{\lambda  z - i(nz - \eta )} \,.
\end{eqnarray}

\end{document}